\documentstyle{livrev}

\RequirePackage{epsf}
\RequirePackage{longtable}

\begin{document}

\title{Perturbative Quantum Gravity and its Relation to
Gauge Theory}

\author{Zvi Bern \\
        Department of Physics and Astronomy\\
        University of California at Los Angeles \\
        Los Angeles, CA 90095 \\
        e-mail:bern@physics.ucla.edu \\
        http://www.physics.ucla.edu/${\sim \atop \null}\hskip -.03 cm$bern\\
\\
\small{(last modified: June, 2002)}
}

\date{}
\maketitle

\begin{abstract}

In this review we describe a non-trivial relationship between
perturbative gauge theory and gravity scattering amplitudes.
At the semi-classical or tree level, the scattering amplitudes of
gravity theories in flat space can be expressed as a sum of products
of well defined pieces of gauge theory amplitudes. These relationships
were first discovered by Kawai, Lewellen, and Tye in the context of
string theory, but hold more generally.  In particular, they hold for
standard Einstein gravity. A method based on $D$-dimensional unitarity
can then be used to systematically construct all quantum loop
corrections order-by-order in perturbation theory using as input the
gravity tree amplitudes expressed in terms of gauge theory ones.  More
generally, the unitarity method provides a means for perturbatively
quantizing massless gravity theories without the usual formal
apparatus associated with the quantization of constrained systems.  As
one application, this method was used to demonstrate that maximally
supersymmetric gravity is less divergent in the ultraviolet than
previously thought.

\end{abstract}


\newpage

%===================================================================================

\section{Introduction}
\label{section:Introduction}

Since its inception, it has been clear that General Relativity has
many striking similarities to gauge theories.  Both are based on the
idea of local symmetry and therefore share a number of formal
properties.  Nevertheless, their dynamical behavior can be quite
different.  While Maxwell electrodynamics describes a long-range force
similar to the situation with gravity, the non-abelian gauge theories
used to describe the weak and strong nuclear forces have rather
different behaviors.  Quantum chromodynamics, which describes the
strong nuclear forces, for example, exhibits confinement of particles
carrying the non-abelian gauge charges. Certainly, there is no obvious
corresponding property for gravity.  Moreover, consistent quantum
gauge theories have existed for a half century, but as yet no
satisfactory quantum field theory of gravity has been constructed;
indeed, there are good arguments suggesting that it is not possible to
do so.  The structures of the Lagrangians are also rather different:
the non-abelian Yang-Mills Lagrangian contains only up to four-point
interactions while the Einstein-Hilbert Lagrangian contains infinitely
many.

Despite these differences, string theory teaches us that gravity and
gauge theories can, in fact, be unified.  The Maldacena
conjecture~\cite{Maldacena98,Aharony00}, for example, relates the weak
coupling limit of a gravity theory on an anti-de Sitter background to
a strong coupling limit of a special supersymmetric gauge field
theory.  There is also a long history of papers noting that gravity
can be expressed as a gauging of Lorentz
symmetry~\cite{Utiyama56,Ivanenko83,Hehl95}, as well as examples
of non-trivial similarities between classical solutions of gravity and
non-abelian gauge theories~\cite{Singleton99}.
In this review a different, but very general, relationship between the
weak coupling limits of both gravity and gauge theories will be
described.  This relationship allows gauge theories to be used
directly as an aid for computations in perturbative quantum gravity.

The relationship discussed here may be understood most easily from
string perturbation theory.  At the semi-classical or ``tree level,''
Kawai, Lewellen and Tye (KLT)~\cite{KLT} derived a precise set of
formulas expressing closed string amplitudes in terms of sums of
products of open string amplitudes.  In the low-energy limit ({\it
i.e.} anywhere well below the string scale of $10^{19}$ GeV) where
string theory effectively reduces to field theory, the KLT relations
necessarily imply that similar relations must exist between amplitudes in
gravity and gauge field theories: at tree-level in field theory, graviton
scattering must be expressible as a sum of products of well defined
pieces of non-abelian gauge theory scattering amplitudes.  Moreover,
using string based rules, four-graviton amplitudes with one quantum
loop in Einstein gravity were obtained in a form in which the integrands
appearing in the expressions were given as products of integrands
appearing in gauge theory~\cite{BDS,DunbarNorridge95}.  These results may
be interpreted heuristically as
\begin{equation}
\hbox{gravity} \sim \hbox{(gauge theory)} \times  \hbox{(gauge theory)}.
\label{HeuristicFormula}
\end{equation}
This remarkable property suggests a much stronger relationship between
gravity and gauge theories than one might have anticipated by
inspecting the respective Lagrangians.

The KLT relations hold at the semi-classical level, {\it i.e.} with no
quantum loops.  In order to exploit the KLT relations in quantum
gravity, one needs to completely reformulate the quantization process;
the standard methods starting either from a Hamiltonian or a
Lagrangian provide no obvious means of exploiting the KLT relations.
There is, however, an alternative approach based on obtaining the
quantum loop contributions directly from the semi-classical tree-level
amplitudes by using $D$-dimensional
unitarity~\cite{Bern94SusyFour,Bern95SusyFour,BernMorgan,Review,Rozowsky}.
These same methods have also been applied to non-trivial calculations
in quantum chromodynamics (see {\it e.g.}
refs.~\cite{BernMorgan,Bern00QCDApplications,Bern02QCDApplications})
 and in supersymmetric gauge
theories (see {\it e.g.}
refs.~\cite{Bern94SusyFour,Bern95SusyFour,BRY,BDDPR}).  In a sense,
they provide a means for obtaining collections of quantum loop-level
Feynman diagrams without direct reference to the underlying Lagrangian
or Hamiltonian.  The only inputs with this method are the
$D$-dimensional tree-level scattering amplitudes.  This makes the
unitarity method ideally suited for exploiting the KLT relations.

An interesting application of this method of perturbatively quantizing
gravity is as a tool for investigating the ultra-violet behavior of
gravity field theories.  Ultraviolet properties are one of the central
issues of perturbative quantum gravity. The conventional wisdom that
quantum field theories of gravity cannot possibly be fundamental
rests on the apparent non-renormalizability of these
theories.  Simple power counting arguments strongly suggest that
Einstein gravity is not renormalizable and therefore can be viewed
only as a low energy effective field theory.  Indeed, explicit
calculations have established that non-supersymmetric theories of
gravity with matter generically diverge at one
loop~\cite{tHooftVeltmanAnnPoin,Deser74,DeserTsao74}, and pure gravity
diverges at two loops~\cite{Goroff86,vandeVen92}.  Supersymmetric
theories are better behaved with the first potential divergence
occurring at three loops~\cite{Deser77,Howe81,Howe89}.  However, no
explicit calculations have as yet been performed to directly
verify the existence of the three-loop supergravity divergences.

The method described here for quantizing gravity is well suited for
addressing the issue of the ultraviolet properties of gravity because
it relates overwhelmingly complicated calculations in quantum gravity
to much simpler (though still complicated) ones in gauge theories.
The first application was for the case of maximally supersymmetric
gravity, which is expected to have the best ultra-violet properties of
any theory of gravity.  This analysis led to the surprising result
that maximally supersymmetric gravity is less divergent~\cite{BDDPR}
than previously believed based on power counting
arguments~\cite{Deser77,Howe81,Howe89}.  This lessening of the power
counting degree of divergence may be interpreted as an additional
symmetry unaccounted for in the original analysis~\cite{Stelle}.  (The
results are inconsistent, however, with an earlier suggestion
\cite{GrisaruSiegel} based on the speculated existence of an
unconstrained covariant off-shell superspace for $N=8$ supergravity,
which in $D=4$ implies finiteness up to seven loops.  The
non-existence of such a superspace was already noted a while
ago~\cite{Howe89}.)  The method also led to the explicit construction
of the two-loop divergence in eleven-dimensional
supergravity~\cite{BDDPR,DeserSeminara99,DeserSeminara00,Bern00Counterterms}.
More recently, it aided the study of divergences in type~I
supergravity theories~\cite{DunbarJulia} where it was noted that they
factorize into products of gauge theory factors.

Other applications include the construction of infinite sequences of
amplitudes in gravity theories.  Given the complexity of gravity
perturbation theory, it is rather surprising that one can obtain
compact expressions for an arbitrary number of external legs, even 
for restricted helicity or spin configurations of the particles.  The
key for this construction is to make use of previously known sequences
in quantum chromodynamics. At tree-level, infinite sequences of
maximally helicity violating amplitudes have been obtained by directly
using the KLT relations~\cite{BGK,Square} and analogous quantum
chromodynamics sequences.  At one loop, by combining the KLT relations
with the unitarity method, additional infinite sequences of gravity
and super-gravity amplitudes have also been
obtained~\cite{AllPlusGrav,MHVGrav}.  They are completely analogous to
and rely on the previously obtained infinite sequences of one-loop
gauge theory
amplitudes~\cite{AllPlusGauge,Bern94SusyFour,Bern95SusyFour}.  These
amplitudes turn out to be also intimately connected to those of
self-dual
Yang-Mills~\cite{Yang77,DuffIsham80,Lenzov87,Lenzov88,Bardeen96,
Cangemi,ChalmersSiegel} and
gravity~\cite{Plebanski75,Duff79,Plebanski96}.  The method has also
been used to explicitly compute two-loop supergravity
amplitudes~\cite{BDDPR} in dimension $D=11$, that were then used to
check M-theory dualities~\cite{GreenTwoLoop}.

Although the KLT relations have been exploited to obtain non-trivial
results in quantum gravity theories, a derivation of these relations
from the Einstein-Hilbert Lagrangian is lacking.  There has, however,
been some progress in this regard.  It turns out that with an
appropriate choice of field variables one can separate the space-time
indices appearing in the Lagrangian into `left' and `right'
classes~\cite{Siegel93A,Siegel93B,Siegel94,BernGrant}, mimicking the similar
separation that occurs in string theory.  Moreover, with further field
redefinitions and a non-linear gauge choice, it is possible to arrange
the off-shell three-graviton vertex so that it is expressible in terms
of a sum of squares of Yang-Mills three-gluon
vertices~\cite{BernGrant}.  It might be possible to extend this 
more generally starting from the formalism of
Siegel~\cite{Siegel93A,Siegel93B,Siegel94}, which contains a complete gravity
Lagrangian with the required factorization of space-time indices.

This review is organized as follows.  In
section~\ref{section:traditional_approach} the Feynman diagram
approach to perturbative quantum gravity is outlined.  The Kawai,
Lewellen and Tye relations between open and closed string tree
amplitudes and their field theory limit are described in
section~\ref{section:KLT_Relations}.  Applications to understanding
and constructing tree-level gravity amplitudes are also described in
this section.  In section~\ref{section:EinsteinHilbert} the
implications for the Einstein-Hilbert Lagrangian are presented. The
procedure for obtaining quantum loop amplitudes from gravity tree
amplitudes is then given in section~\ref{section:trees_to_loops}.  The
application of this method to obtain quantum gravity loop amplitudes
is described in section~\ref{section:gravity_loops}. In
section~\ref{section:divergence_properties} the quantum divergence
properties of maximally supersymmetric supergravity obtained from this
method are described.  The conclusions are found in
section~\ref{section:conclusions}.

There are a number of excellent sources for various subtopics
described in this review.  For a recent review of the status of
quantum gravity the reader may consult the article by
Carlip~\cite{Carlip}. The conventional Feynman diagram approach to
quantum gravity can be found in the Les Houches lectures of
Veltman~\cite{VeltmanGravity}.  A review article containing an early
version of the method described here of using unitarity to construct
complete loop amplitudes is ref.~\cite{Review}.  Excellent reviews
containing the quantum chromodynamics amplitudes used to obtain
corresponding gravity amplitudes are the ones by Mangano and
Parke~\cite{ManganoReview} and by Lance Dixon~\cite{TasiLance}.  These
reviews also provide a good description of helicity techniques which
are extremely useful for explicitly constructing scattering amplitude
in gravity and gauge theories.  Broader textbooks describing quantum
chromodynamics are Refs.~\cite{Peskin95,Weinberg95,Ellis96}.  Chapter
7 of {\it Superstring Theory} by Green, Schwarz, and Witten~\cite{GSW}
contains an illuminating discussion of the relationship of closed and
open string tree amplitudes, especially at the four-point level.  A
somewhat more modern description of string theory may be found in the
book by Polchinski~\cite{Polchinski98A,Polchinski98B}.  Applications
of string methods to quantum field theory are described in a 
recent review by Schubert~\cite{Schubert01}.

\newpage
%=======================================================================

\section{Outline of Traditional Approach to Perturbative Quantum Gravity}
\label{section:traditional_approach}

\subsection{Overview of Gravity Feynman Rules}
\label{subsection:gravity_rules}

Scattering of gravitons in flat space may be described using Feynman
diagrams~\cite{DeWitt67A,DeWitt67B,VeltmanGravity}.  The Feynman rules
for constructing the diagrams are obtained from the Einstein-Hilbert
Lagrangian coupled to matter using standard procedures of quantum
field theory. (The reader may consult any of the textbooks on quantum
field theory~\cite{Peskin95,Weinberg95} for a derivation of the
Feynman rules starting from a given Lagrangian.) For a good source
describing the Feynman rules of gravity, the reader may consult the
classic lectures of Veltman~\cite{VeltmanGravity}.

%FIGURE
%%%%%%%%%%%%%%%%%%%%%%%%%%% 
\begin{figure}[h]
  \def\epsfsize#1#2{1.0#1}
  \centerline{\epsfbox{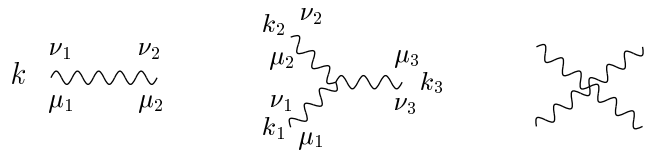}}
  \caption{\it The Feynman propagator and three- and 
   four-point vertices in Einstein gravity.}
  \label{figure:GravityFeynman}
\end{figure}
%%%%%%%%%%%%%%%%%%%%%%%%%%

The momentum-space Feynman rules are expressed in terms of vertices
and propagators as depicted in Fig.~\ref{figure:GravityFeynman}.  In
the figure, space-time indices are denoted by $\mu_i$ and $\nu_i$
while the momenta are denoted by $k$ or $k_i$. In contrast to gauge
theory, gravity has an infinite set of ever more complicated
interaction vertices; the three- and four-point ones are displayed in
the figure.  The diagrams for describing scattering of gravitons from
each other are built out of these propagators and vertices.  Other 
particles can be included in this framework by adding new propagators
and vertices associated with each particle type.
(For the case of fermions coupled to gravity the Lagrangian needs to be
expressed in terms of the vierbein instead of the metric before the
Feynman rules can be constructed.)

According to the Feynman rules, each leg or vertex represents 
a specific algebraic expression depending on the choice of field
variables and gauges. For example, the graviton Feynman propagator in the
commonly used De~Donder gauge is:
\begin{equation}
 P_{\mu_1 \nu_1 ; \mu_2 \nu_2} = 
 {1 \over 2} 
\Bigl[\eta_{\mu_1\mu_2} \eta_{\mu_2 \nu_2} 
+ \eta_{\mu_1\nu_2} \eta_{\nu_1\mu_2} 
- {2\over D-2} \eta_{\mu_1\nu_1} \eta_{\mu_2 \nu_2} \Bigr] 
{i\over k^2 + i\epsilon}\,. 
\label{GravityPropagator}
\end{equation}
The three-vertex is much more complicated and the expressions
may be found in DeWitt's articles~\cite{DeWitt67A,DeWitt67B} or in 
Veltman's lectures~\cite{VeltmanGravity} .  For simplicity,
only a few of the terms of the three-vertex are displayed: 
\arraycolsep 0.14 em
\begin{eqnarray}
G_{\rm De\ Donder}^{\mu_1 \nu_1,\mu_2\nu_2,\mu_3\nu_3}(k_1,k_2,k_3)\ &\sim&
k_1 \cdot k_2 \eta^{\mu_1\nu_1}\eta^{\mu_2\nu_2}\eta^{\mu_3\nu_3} +
k_1^{\mu_3}k_2^{\nu_3}\eta^{\mu_1\mu_2}\eta^{\nu_1\nu_2} \nonumber\\
&& \hskip 2 cm 
\ +\ \hbox{many other terms}\,
\label{ThreeDeDonder}
\end{eqnarray}
where the indices associated with each graviton are depicted in the
three-vertex of Fig.\ref{figure:GravityFeynman}, {\it i.e.}, the two
indices of graviton $i=1,2,3$ are $\mu_i \nu_i$.

%FIGURE
%%%%%%%%%%%%%%%%%%%%%%%%%%% 
\begin{figure}[h]
  \def\epsfsize#1#2{.5#1}
  \centerline{\epsfbox{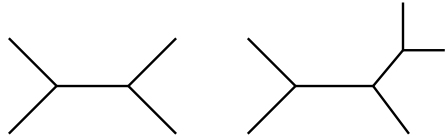}}
  \caption{\it Sample gravity tree-level Feynman diagrams. The lines 
  represent any particles in a gravity theory.}
  \label{figure:GravityTrees}
\end{figure}
%%%%%%%%%%%%%%%%%%%%%%%%%%

%FIGURE
%%%%%%%%%%%%%%%%%%%%%%%%%%% 
\begin{figure}[h]
  \def\epsfsize#1#2{.5#1}
  \centerline{\epsfbox{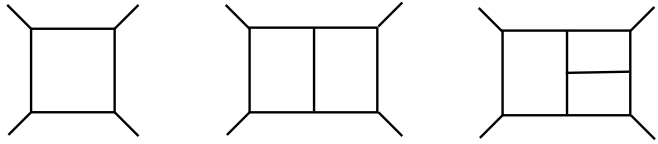}}
  \caption{\it Sample loop-level Feynman diagrams. Each additional loop 
      represents an extra power of Planck's constant.}
  \label{figure:GravityLoops}
\end{figure}
%%%%%%%%%%%%%%%%%%%%%%%%%%

The loop expansion of Feynman diagrams provide a systematic quantum
mechanical expansion in Planck's constant $\hbar$.  The tree-level
diagrams such as those in Fig.~\ref{figure:GravityTrees} are
interpreted as (semi) classical scattering processes while the
diagrams with loops are the true quantum mechanical effects: each loop
carries with it a power of $\hbar$.  According to the Feynman rules,
each loop represents an integral over the momenta of the intermediate
particles.  The behavior of these loop integrals is the key for
understanding the divergences of quantum gravity.

\subsection{Divergences in Quantum Gravity}

In general, the loop momentum integrals in a quantum field theory will
diverge in the ultraviolet where the momenta in the loops become
arbitrarily large.  Unless these divergences are of the right form
they indicate that a theory cannot be interpreted as 
fundamental, but is instead valid only at low energies.  Gauge
theories such as quantum chromodynamics are renormalizable:
divergences from high energy scales can be absorbed into redefinitions
of the original parameters appearing in the theory. In quantum gravity,
on the other hand, it is not possible to re-absorb divergences in the
original Lagrangian for a very simple reason: the gravity coupling $
\kappa = \sqrt{32 \pi G_N}$, where $G_N$ is Newton's constant, carries
dimensions of length (in units where $\hbar = c = 1$). By dimensional 
analysis, any divergence must be proportional to terms with extra
derivatives compared to the original Lagrangian and are thus of a
different form.  This may be contrasted to the gauge theory situation
where the coupling constant is dimensionless, allowing for the theory
to be renormalizable.

The problem of non-renormalizability of quantum gravity does not mean
that quantum mechanics is incompatible with gravity, only that quantum
gravity should be treated as an effective field
theory~\cite{Weinberg79,Gasser85,Donoghue94,Kaplan95,Manohar96} for
energies well below the Planck scale of $10^{19}$ GeV (which is, of
course, many orders of magnitude beyond the reach of any conceivable
experiment).  In an effective field theory, as one computes higher loop
orders new and usually unknown couplings need to be introduced to
absorb the divergences.  Generally, these new couplings are suppressed
at low energies by ratios of energy to the fundamental high energy
scale, but at sufficiently high energies the theory loses its
predictive power.  In quantum gravity this happens at the Planck
scale.

Quantum gravity based on the Feynman diagram expansion allows for a
direct investigation of the non-renormalizability issue.  For a theory
of pure gravity with no matter, amazingly, the one-loop divergences
cancel, as demonstrated by 't~Hooft and Veltman\footnote{This happens
because field redefinitions exist that can be used to remove the
potential divergences.}~\cite{tHooftVeltmanAnnPoin}. Unfortunately,
this result is ``accidental,'' since it does not hold generically when
matter is added to the theory or when the number of loops is
increased.  Explicit calculations have shown that
non-supersymmetric theories of gravity with matter generically diverge
at one loop~\cite{tHooftVeltmanAnnPoin,Deser74,DeserTsao74}, and pure
gravity diverges at two loops~\cite{Goroff86,vandeVen92}.  The
two-loop calculations were performed using various improvements to the
Feynman rules such as the background field
method~\cite{tHooft75,DeWitt81,Background}.  

Supersymmetric theories of gravity are known to have less severe
divergences. In particular, in any four-dimensional supergravity
theory, supersymmetry Ward
identities~\cite{Grisaru77SWIA,Grisaru77SWIB} forbid all possible
one-loop~\cite{OneLoopSUGRA} and two-loop~\cite{Grisaru77,Tomboulis77}
divergences.  There is a candidate divergence at three loops for all
supergravities including the maximally extended $N=8$
version~\cite{Deser77,Howe81,KalloshNeight,Howe89}.  However, no
explicit three-loop (super) gravity calculations have been performed
to confirm the divergence.  In principle it is possible that the coefficient
of a potential divergence obtained by power counting can vanish,
especially if the full symmetry of the theory is taken into account.
As described in section~\ref{section:divergence_properties}, this is
precisely what does appear to happen~\cite{BDDPR,Stelle} in the case
of maximally supersymmetric supergravity.

\begin{figure}[h]
  \def\epsfsize#1#2{0.3#1}
  \centerline{\epsfbox{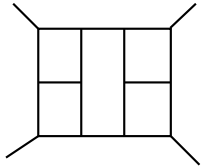}}
  \caption{\it An example of a five-loop diagram.}
  \label{figure:Multiloop}
\end{figure}

The reason no direct calculation of the three-loop supergravity
divergences has been performed is the overwhelming technical
difficulties associated with multi-loop gravity Feynman diagrams.  In
multi-loop calculations the number of algebraic terms proliferates
rapidly beyond the point where computations are practical.  As a
particularly striking example, consider the five-loop diagram in
Fig.~\ref{figure:Multiloop}, which, as noted in
section~\ref{section:divergence_properties}, is of interest for
ultraviolet divergences in maximal $N=8$ supergravity in $D=4$.  In
the standard De~Donder gauge this diagram contains twelve vertices,
each of the order of a hundred terms, and sixteen graviton
propagators, each with three terms, for a total of roughly $10^{30}$
terms, even before having evaluated any integrals.  This is obviously
well beyond what can be implemented on any computer. The standard
methods for simplifying diagrams, such as background-field gauges and
superspace, are unfortunately insufficient to reduce the problem to
anything close to manageable levels. The alternative of using string-
based methods that have proven to be useful at one loop and in
certain two-loop
calculations~\cite{Long,BDS,Schmidt94,DunbarNorridge95,DunbarNorridge97,
Dunbar02,Schubert01} also does not as yet provide a practical means
for performing multi-loop scattering amplitude
calculations~\cite{Roland92,Roland96,DiVecchia96,Roland98,Frizzo00},
especially in gravity theories.

%FIGURE
%%%%%%%%%%%%%%%%%%%%%%%%%%% 
%%%%%%%%%%%%%%%%%%%%%%%%%%

\subsection{Gravity and Gauge Theory Feynman Rules}

\begin{figure}[h]
  \def\epsfsize#1#2{0.4#1}
  \centerline{\epsfbox{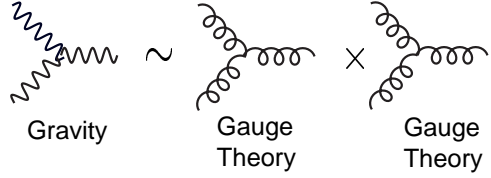}}
  \caption{\it String theory suggests that the three-graviton vertex can
be expressed in terms of products of three-gluon vertices.}
  \label{figure:ThreeVertex}
\end{figure}

The heuristic relation (\ref{HeuristicFormula}) suggests a possible
way to deal with multi-loop diagrams such as the one in
Fig.~\ref{figure:Multiloop} by somehow factorizing gravity amplitudes
into products of gauge theory ones.  Since gauge theory Feynman rules
are inherently much simpler than gravity Feynman rules, it clearly
would be advantageous to re-express gravity perturbative expansions in
terms of gauge theory ones.  As a first step, one might, for example,
attempt to express the three-graviton vertex as a product of two
Yang-Mills vertices, as depicted in Fig.~\ref{figure:ThreeVertex}:
\begin{equation}
G_{\rm factorizing}^{\mu_1 \nu_1,\mu_2\nu_2,\mu_3\nu_3}(k_1,k_2,k_3)\ \sim\
V_{\rm YM}^{\mu_1 \mu_2 \mu_3}(k_1,k_2,k_3)
V_{\rm YM}^{\nu_1 \nu_2 \nu_3}(k_1,k_2,k_3)
\label{ThreeFactorizing}
\end{equation}
where the two indices of each graviton labeled by $i=1,2,3$ are 
$\mu_i \nu_i$, {\it i.e.} $h_{\mu_i \nu_i}$.

% FIGURE
%%%%%%%%%%%%%%%%%%%%%%%%%%% 
%%%%%%%%%%%%%%%%%%%%%%%%%%

Such relations, however, do not hold in any of the standard formulations
of gravity.  For example, the three vertex in the standard De~Donder
gauge~(\ref{ThreeDeDonder}) contains traces over gravitons, {\it i.e.} a
contraction of indices of a single graviton.  For physical gravitons
the traces vanish, but for gravitons appearing inside Feynman diagrams
it is in general crucial to keep such terms.  A necessary condition
for obtaining a factorizing three-graviton
vertex~(\ref{ThreeFactorizing}) is that the ``left'' $\mu_i$ indices
never contract with the ``right'' $\nu_i$ indices.  This is clearly
violated by the three-vertex in Eq.~(\ref{ThreeDeDonder}).  Indeed,
the standard formulations of quantum gravity generate a plethora of
terms that violate the heuristic relation~(\ref{HeuristicFormula}).

In section~\ref{section:EinsteinHilbert} the question of how one
rearranges the Einstein action to be compatible with string theory
intuition is returned to.  However, in order to give a precise meaning
to the heuristic formula~(\ref{HeuristicFormula}) and to demonstrate
that scattering amplitudes in gravity theories can indeed be
obtained from standard gauge theory ones, a completely different
approach from the standard Lagrangian or Hamiltonian ones is required.
This different approach is described in the next section.

\newpage
%=============================================================================

\section{The Kawai-Lewellen-Tye Relations}
\label{section:KLT_Relations}

Our starting point for constructing perturbative quantum gravity is
the Kawai, Lewellen, and Tye (KLT) relations~\cite{KLT} between closed
and open string tree-level amplitudes.  Since closed string theories are
theories of gravity and open string theories include gauge bosons, in
the low energy limit, where string theory reduces to field theory,
these relations then necessarily imply relations between gravity and
gauge theories. The realization that ordinary gauge and gravity
field theories emerge from the low energy limit of string theories 
has been appreciated for nearly three decades. (See, for example,
Refs.~\cite{Yoneya74,Scherk,Sannan86,GSB,GSW,Polchinski98A,Polchinski98B}).

%%%%%
\subsection{The KLT Relations in String Theory}

The KLT relations between open and closed string theory amplitudes
can be motivated by the observation is that any
closed string vertex operator for the emission of a closed string
state (such as a graviton) is a product of open string vertex
operators (see {\it e.g.} Ref.~\cite{GSW}),
\begin{equation}
V^{\rm closed} = V_{\rm left}^{\rm open}\, \times \,  
\overline{V}_{\rm right}^{\rm  open} \,.
\label{ClosedVertex}
\end{equation}
This product structure is then reflected in the amplitudes.  Indeed,
the celebrated Koba-Nielsen form of string
amplitudes~\cite{KobaNielsen}, which may be obtained by evaluating
correlations of the vertex operators, factorize at the level of the
integrands before world sheet integrations are performed.  Amazingly,
Kawai, Lewellen, and Tye were able to demonstrate a much stronger
factorization: complete closed string amplitudes factorize into
products of open string amplitudes, even {\it after} integration over
the world sheet variables.  (A description of string theory scattering
amplitudes and the history of their construction may be found in
standard books on string theory~\cite{GSW,Polchinski98A,Polchinski98B}.)

As a simple example of the factorization property of string 
theory amplitudes, the four-point partial amplitude of open 
superstring theory for scattering any of the massless modes is given by 
\begin{equation}
A_4^{\rm open} = 
-{1\over 2} g^2 {\Gamma(\alpha's)  \Gamma(\alpha' t) \over 
\Gamma(1 + \alpha's + \alpha' t)} 
\xi^A \xi^B \xi^C \xi^D
K_{A B C D}(k_1, k_2, k_3, k_4) \,,
\label{OpenStringFourPoint}
\end{equation}
where $\alpha'$ is the open string Regge slope proportional to the
inverse string tension, $g$ is the gauge theory coupling, and $K$ is a
gauge invariant kinematic coefficient depending on the momenta $k_1,
\ldots, k_4$. Explicit forms of $K$ may be found in Ref.~\cite{GSW}.
(The metric is taken here to have signature
$(+,-,-,-)$.)  In this and subsequent expressions, $s = (k_1 +
k_2)^2$, $t = (k_1 + k_2)^2$ and $u = (k_1 + k_3)^2$.  The indices can
be either vector, spinor or group theory indices and the $\xi^A$ can
be vector polarizations, spinors, or group theory matrices, depending
on the particle type.  These amplitudes are the open string partial
amplitudes before they are dressed with Chan-Paton~\cite{ChanPaton} group
theory factors and summed over non-cyclic permutations to form
complete amplitudes. (Any group theory indices in
Eq.~(\ref{OpenStringFourPoint}) are associated with string world sheet
charges arising from possible compactifications.)  For the case of a
vector, $\xi^A$ is the usual polarization vector.  Similarly, the
four-point amplitudes corresponding to a heterotic closed
superstring~\cite{Gross85Heterotic,Gross86Heterotic} are,
\arraycolsep 0.14 em
\begin{eqnarray}
M_4^{\rm closed} &=& 
\kappa^2  \sin\Bigl({\alpha' \pi t \over 4} \Bigr) \times 
 {\Gamma(\alpha's/4)  \Gamma(\alpha't/4) \over 
      \Gamma(1 + \alpha's/4 + \alpha' t/4)}
\times  {\Gamma(\alpha't/4)  \Gamma(\alpha' u/4) 
      \over \Gamma(1 + \alpha't/4 + \alpha'u/4)} \nonumber \\
&& \times 
\xi^{AA'} \xi^{BB'} \xi^{CC'} \xi^{DD'}
K_{A B C D}(k_1/2, k_2/2, k_3/2, k_4/2) \nonumber \\
&& \hskip 3 cm \times
K_{A' B' C' D'}(k_1/2, k_2/2, k_3/2, k_4/2) \,,
\label{ClosedStringFourPoint}
\end{eqnarray}
where $\alpha'$ is the open string Regge slope or equivalently twice
the close string one.
Up to prefactors, the
replacements $\xi^A \xi^{A'} \rightarrow \xi^{A A'}$ and substituting  $k_i
\rightarrow k_i/2$, the closed string
amplitude~(\ref{ClosedStringFourPoint}) is a product of the open
string partial amplitudes~(\ref{OpenStringFourPoint}).  For the case of
external gravitons the $\xi^{AA'}$ are ordinary graviton polarization
tensors.  For further reading, Chapter 7 of {\it Superstring Theory}
by Green, Schwarz, and Witten~\cite{GSW} provides an especially
enlightening discussion of the four-point amplitudes in various string
constructions.

As demonstrated by KLT, the property that closed string tree amplitudes
can be expressed in terms of products of open string tree amplitudes
is completely general for any string states and for any number of
external legs. In general, it holds also for each of the huge number
of possible string
compactifications~\cite{Narain86,Narain87,Dixon85,Dixon86,Kawai87,
Antoniadis87}.

 An essential part of the factorization of the amplitudes
is that any closed-string state is a direct product of two open-string
states.  This property directly follows from the factorization of the
closed-string vertex operators (\ref{ClosedVertex}) into products of open-
string vertex operators.  In general, for every closed-string
state there is a Fock space decomposition
\begin{equation}
\vert \hbox{closed string state} \rangle
\ =\ \vert  \hbox{open string state} \rangle 
\otimes \vert\hbox{open string state} \rangle \,.
\end{equation}
In the low energy limit this implies that states in a gravity field
theory obey a similar factorization, 
\begin{equation}
\vert \hbox{gravity theory state} \rangle
\ =\ \vert  \hbox{gauge theory state} \rangle 
\otimes \vert\hbox{gauge theory state} \rangle \,.
\label{StateFactorization}
\end{equation}
For example, in four dimensions each of the two physical helicity
states of the graviton are given by the direct product of two vector
boson states of identical helicity. The cases where the vectors have
opposite helicity correspond to the antisymmetric tensor and dilaton.
Similarly, a spin 3/2 gravitino state, for example, is a direct product
of a spin 1 vector and spin 1/2 fermion.  Note that decompositions of
this type are not especially profound for free field theory and
amount to little more than decomposing higher spin states as direct
products of lower spin ones.  What is profound is that the
factorization holds for the full non-linear theory of gravity.

%%%%%
\subsection{The KLT Relations in Field Theory}

The fact that the KLT relations hold for the extensive
variety of compactified string models~\cite{Narain86,Narain87,
Dixon85,Dixon86,Kawai87,Antoniadis87} implies that they should
also be generally true in field theories of gravity.
For the cases of four- and five-particle scattering amplitudes, in the
field theory limit the KLT relations~\cite{KLT} reduce to:
\begin{equation}
M_4^{\rm tree}  (1,2,3,4) = 
 - i s_{12} A_4^{\rm tree} (1,2,3,4) \, A_4^{\rm tree}(1,2,4,3)\,,
\label{KLTFourPoint}
\end{equation}
\arraycolsep 0.14 em
\begin{eqnarray}
M_5^{\rm tree}(1,2,3,4,5) & = &
i s_{12} s_{34}  A_5^{\rm tree}(1,2,3,4,5)
                                 A_5^{\rm tree}(2,1,4,3,5) \nonumber\\
&& \hskip .3 cm \null
 + i s_{13}s_{24} A_5^{\rm tree}(1,3,2,4,5) \, A_5^{\rm tree}(3,1,4,2,5)\,,
\label{KLTFivePoint}
\end{eqnarray}
where the $M_n$'s are tree-level amplitudes in a gravity theory, the
$A_n$'s are color-stripped tree-level amplitudes in a gauge theory and
$s_{ij}\equiv (k_i+k_j)^2$. In these equations the polarization and
momentum labels are suppressed, but the label ``$j= 1,\ldots, n$'' is
kept to distinguish the external legs.  The coupling constants have
been removed from the amplitudes, but are reinserted below in
Eqs.~(\ref{FullGaugeTheory}) and (\ref{GravityCouplingAmplitude}).  An
explicit generalization to $n$-point field theory gravity amplitudes
may be found in appendix~A of Ref.~\cite{MHVGrav}.  The KLT relations
before the field theory limit is taken may, of course, be found in the
original paper~\cite{KLT}.

The KLT equations generically hold for any closed string states, using
their Fock space factorization into pairs of open string states.
Although not obvious, the gravity amplitudes~(\ref{KLTFourPoint}) and
(\ref{KLTFivePoint}) have all the required symmetry under interchanges
of identical particles.  (This is easiest to demonstrate in string
theory by making use of an $SL(2,Z)$ symmetry on the string world
sheet.)  

In the field theory limit the KLT equations must hold in any
dimension, because the gauge theory amplitudes appearing on the
right-hand-side have no explicit dependence on the space-time
dimension; the only dependence is implicit in the number of components
of momenta or polarizations.  Moreover, if the equations hold in, say,
ten dimensions, they must also hold in all lower dimensions since one
can truncate the theory to a lower dimensional subspace.

The amplitudes on the left-hand side of Eqs.~(\ref{KLTFourPoint}) and
(\ref{KLTFivePoint}) are exactly the scattering amplitudes that one
obtains via standard gravity Feynman
rules~\cite{DeWitt67A,DeWitt67B,VeltmanGravity}.  The gauge theory
amplitudes on the right-hand-side may be computed via standard Feynman
rules available in any modern textbook on quantum field
theory~\cite{Peskin95,Weinberg95}.  After computing the full gauge
theory amplitude, the color-stripped partial amplitudes $A_n$
appearing in the KLT relations~(\ref{KLTFourPoint})
and~(\ref{KLTFivePoint}), may then be obtained by expressing the
full amplitudes in a color trace
basis~\cite{Berends87,Kosower88,Mangano88,ManganoReview,TasiLance}:
\begin{equation}
{\cal A}_n^{\rm tree} (1,2,\ldots n) = 
 g^{(n-2)}  \sum_{\sigma } 
{\rm Tr}\left( T^{a_{\sigma(1)}} 
\cdots  T^{a_{\sigma(n)}} \right)
 A_n^{\rm tree}(\sigma(1), \ldots, \sigma(n)) \,, \hskip .3 cm 
\label{FullGaugeTheory}
\end{equation}
where the sum runs over the set of all permutations, but with cyclic
rotations removed and $g$ as the gauge theory coupling constant.  The
$A_n$ partial amplitudes that appear in the KLT relations are defined
as the coefficients of each of the independent color traces. In this
formula, the $T^{a_i}$ are fundamental representation matrices for the
Yang-Mills gauge group $SU(N_c)$, normalized so that ${\rm Tr}(T^aT^b)
= \delta^{ab}$. Note that the $A_n$ are completely independent of the
color and are the same for any value of $N_c$.
Eq.~(\ref{FullGaugeTheory}) is quite similar to the way full open
string amplitudes are expressed in terms of the string partial
amplitudes by dressing them with Chan-Paton color
factors~\cite{ChanPaton}.

Instead, it is somewhat more convenient to use color-ordered Feynman
rules~\cite{ManganoReview,TasiLance,Review} since they directly give
the $A_n$ color-stripped gauge theory amplitudes appearing in the KLT
equations.  These Feynman rules are depicted in
Fig.~\ref{figure:Rules}.  When obtaining the partial amplitudes from
these Feynman rules it is essential to order the external legs
following the order appearing in the corresponding color trace.
One may view the color-ordered gauge theory rules as a new set of
Feynman rules for gravity theories at tree level, since the KLT
relations allow one to convert the obtained diagrams to tree-level
gravity amplitudes~\cite{Square} as follows:

% FIGURE
%%%%%%%%%%%%%%%%%%%%%%%%%%% 
\begin{figure}[h]
  \def\epsfsize#1#2{0.8#1} \centerline{\epsfbox{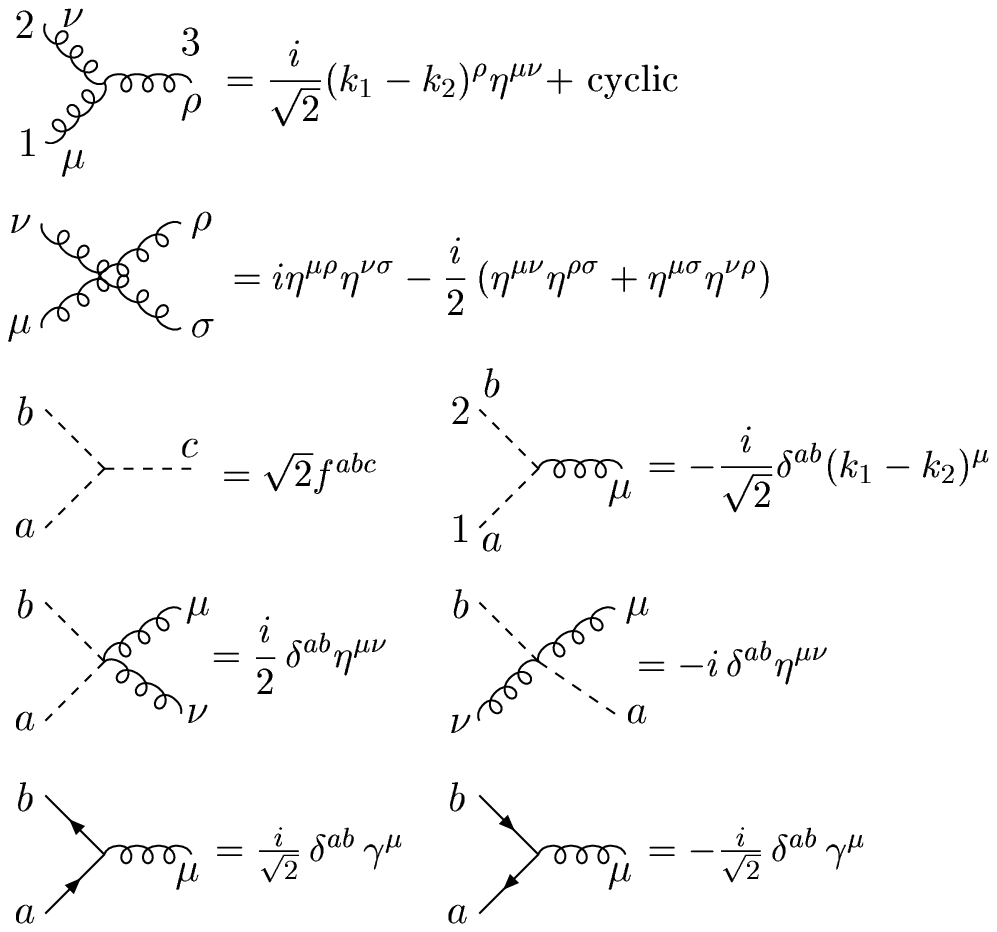}}
  \caption{\it The color-ordered gauge theory Feynman rules for
  obtaining tree-level scattering amplitudes for gravity coupled to matter, via
  the KLT equations. The Greek indices are space-time ones and the
  Latin ones are group theory ones.  The curly lines are vectors,
  dotted ones scalars, and the solid ones fermions.}
  \label{figure:Rules}
\end{figure}
%%%%%%%%%%%%%%%%%%%%%%%%%%

To obtain the full amplitudes from the KLT relations in
Eqs.~(\ref{KLTFourPoint}), (\ref{KLTFivePoint}) and their $n$-point
generalization, the couplings need to be reinserted.  In particular,
when all states couple gravitationally, the full gravity amplitudes
including the gravitational coupling constant are:
\begin{equation}
{\cal M}_n^{\rm tree} (1,\ldots n) = 
\left({  \kappa \over 2} \right)^{(n-2)} 
M_n^{\rm tree}(1,\ldots n)\,, \hskip 1 cm 
\label{GravityCouplingAmplitude}
\end{equation}
%2
where $\kappa^2 = 32\pi G_N$ expresses the coupling $\kappa$ in terms
of Newton's constant $G_N$.  In general, the precise combination of
coupling constants depends on how many of the interactions are gauge 
or other interactions and how many are gravitational.

For the case of four space-time dimensions, it is very convenient to
use helicity representation for the physical
states~\cite{Causmaecker81,Kleiss85,Xu87}.  With helicity amplitudes
the scattering amplitudes in either gauge or gravity theories are, in
general, remarkably compact, when compared with expressions where formal
polarization vectors or tensors are used.  For each helicity, the
graviton polarization tensors satisfy a simple relation to gluon
polarization vectors:
\begin{equation}
\varepsilon_{\mu\nu}^+ = \varepsilon_{\mu}^+ \varepsilon_{\nu}^+\,,
\hskip 2 cm 
\varepsilon_{\mu\nu}^- = \varepsilon_{\mu}^- \varepsilon_{\nu}^- \,.
\end{equation}
The $\varepsilon_{\mu}^\pm$ are essentially ordinary circular
polarization vectors associated with, for example, circularly
polarized light.  The graviton polarization tensors defined in this
way automatically are traceless, $\varepsilon_{\mu}^{\pm\mu} = 0$,
because the gluon helicity polarization vectors satisfy
$\varepsilon^\pm \cdot \varepsilon^\pm = 0$.  They are also
transverse, $\varepsilon_{\mu\nu}^\pm k^\nu = \varepsilon_{\mu\nu}^\pm
k^\mu = 0$, because the gluon polarization vectors satisfy
$\varepsilon^\pm \cdot k = 0$, where $k^\mu$ is the four momentum of
either the graviton or gluon.

\subsection{Tree-level Applications}

Using a helicity representation~\cite{Causmaecker81,Kleiss85,Xu87},
Berends, Giele, and Kuijf (BGK)~\cite{BGK} were the first to exploit
the KLT relations to obtain amplitudes in Einstein gravity.  In
quantum chromodynamics (QCD) an infinite set of helicity amplitudes
known as the Parke-Taylor
amplitudes~\cite{ParkeTaylor86,BerendsGiele88,Kosower90LightCone} were
already known.  These maximally helicity violating (MHV) amplitudes
describe the tree-level scattering of $n$ gluons when all gluons but
two have the same helicity, treating all particles as outgoing.  (The
tree amplitudes in which all or all but one of the helicities are
identical vanish.)  BGK used the KLT relations to directly obtain
graviton amplitudes in pure Einstein gravity, using the known QCD
results as input.  Remarkably, they also were able to obtain a compact
formula for $n$-graviton scattering with the special helicity
configuration in which two legs are of opposite helicity from the
remaining ones.

\begin{figure}[h]
  \def\epsfsize#1#2{0.9#1} \centerline{\epsfbox{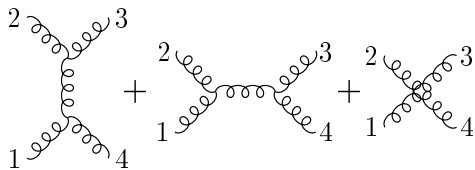}}
  \caption{\it The three color-ordered Feynman diagrams contributing to 
         the QCD partial amplitude in Eq.~(\ref{FourGluonAmplitude}).}
  \label{figure:Gluon}
\end{figure}

As a particularly simple example, the color-stripped four-gluon tree 
amplitude with two minus helicities and two positive helicities
in QCD is given by
\begin{equation}
A_4^{\rm tree}(1^-_g, 2^-_g, 3^+_g, 4^+_g) =  {s_{12} \over s_{23}} \,,
\label{FourGluonAmplitude}
\end{equation}
\begin{equation}
A_4^{\rm tree}(1^-_g, 2^-_g, 4^+_g, 3^+_g) =  {s_{12} \over s_{24}} \,,
\label{FourGluonAmplitudeB}
\end{equation}
where the $g$ subscripts signify that the legs are gluons and the
$\pm$ superscripts signify the helicities.  With the conventions used
here, helicities are assigned by treating all particles as
outgoing.  (This differs from another common choice which is to keep
track of which particles are incoming and which are outgoing.)  In
these amplitudes, for simplicity, overall phases have been  removed.  The
gauge theory partial amplitude in Eq.~(\ref{FourGluonAmplitude}) may
be computed using the color-ordered Feynman diagrams depicted in
Fig.~\ref{figure:Gluon}.  The diagrams for the partial amplitude in
Eq.~(\ref{FourGluonAmplitudeB}) are similar except that the labels for
legs 3 and 4 are interchanged. Although QCD contains fermion quarks,
they do not contribute to tree amplitudes with only external gluon legs
because of fermion number conservation; for these amplitudes QCD is
entirely equivalent to pure Yang-Mills theory.

% FIGURE
%%%%%%%%%%%%%%%%%%%%%%%%%%% 
%%%%%%%%%%%%%%%%%%%%%%%%%%

The corresponding four-graviton amplitude follows 
from the KLT equation~(\ref{KLTFourPoint}).  After including the
coupling from Eq.~(\ref{GravityCouplingAmplitude}), the four-graviton
amplitude is:
\begin{equation}
{\cal M}_4^{\rm tree} (1^-_h, 2^-_h, 3^+_h, 4^+_h) = 
\left({\kappa \over 2} \right)^{2}  s_{12} \times  {s_{12} \over s_{23}}
\times  {s_{12} \over s_{24}} \,,
\label{GravityExample}
\end{equation}
where the subscript $h$ signifies that the particles are gravitons and,
as with the gluon amplitudes, overall phases are removed. As for the
case of gluons, the $\pm$ superscripts signify the helicity of the
graviton. This amplitude necessarily must be identical to the result
for pure Einstein gravity with no other fields present, because any
other states, such as an anti-symmetric tensor, dilaton, or fermion, do
not contribute to $n$-graviton tree amplitudes. The reason is similar
to the reason why the quarks do not contribute to pure glue tree amplitudes in
QCD. These other physical states contribute only when they appear as
an external state, because they couple only in pairs to the graviton.
Indeed, the amplitude (\ref{GravityExample}) is in complete agreement
with the result for this helicity amplitude obtained by direct
diagrammatic calculation using the pure gravity Einstein-Hilbert
action as the starting point~\cite{Berends74} (and taking into account the
different conventions for helicity).

The KLT relations are not limited to pure gravity
amplitudes.  Cases of gauge theory coupled to gravity have
also been discussed in Ref.~\cite{Square}.
For example, by applying the Feynman rules in Fig.~\ref{figure:Rules}, 
one can obtain amplitudes for gluon amplitudes dressed with gravitons.
A sampling of these, to leading order in the graviton coupling, 
is:
\begin{eqnarray}
&& M_4(1_g^+,2_g^+, 3_h^-, 4_h^-) =  0\,, \nonumber \\
&& M_4(1_g^-, 2_g^+, 3_h^-, 4_h^+) =  \Bigl({\kappa\over 2} \Bigr)^2
        \biggl| {s_{23}^5
                     \over s_{13} s_{12}^2} \biggr|^{1/2} \,, \nonumber\\
&&M_5(1_g^+, 2_g^+, 3_g^+, 4_g^-, 5_h^-) =  g^2 \Bigl({\kappa\over 2} \Bigr)
   \biggl| {s_{45}^4 \over s_{12} s_{23} s_{34} s_{41} } 
    \biggr|^{1/2}\,, \nonumber \\
&&M_5(1_g^+, 2_g^+, 3_h^+, 4_h^-, 5_h^-) = 0 \,, \nonumber \\
\end{eqnarray}
for the coefficients of the color traces ${\rm Tr} [T^{a_1} \cdots
T^{a_m}]$ following the ordering of the gluon legs.  Again, for
simplicity, overall phases are eliminated from the amplitudes.  (In
Ref.~\cite{Square} mixed graviton matter amplitudes including the
phases may be found.)

These formulae have been generalized to infinite sequences of
maximally helicity-violating tree amplitudes for gluon amplitudes
dressed by external gravitons.  The first of these were obtained by
Selivanov using a generating function technique~\cite{Selivanov}.
Another set was obtained using the KLT relations to find the pattern
for an arbitrary number of legs~\cite{Square}.  In doing this, it is
extremely helpful to make use of the analytic properties of amplitudes
as the momenta of various external legs become soft ({\it i.e.} $k_i
\rightarrow 0$) or collinear ({\it i.e.}  $k_i$ parallel to $k_j$), as discussed
in the next subsection.

Cases involving fermions have not been systematically studied, but at
least for the case with a single fermion pair the KLT equations
can be directly applied using the Feynman rules in
Fig.~\ref{figure:Rules}, without any modifications.  For example, in a
supergravity theory, the scattering of a gravitino by a graviton is
\begin{eqnarray}
{\cal M}_4^{\rm tree} (1^-_{\tilde h}, 2^-_{h {\vphantom{\tilde h}}}, 
                   3^+_{h {\vphantom{\tilde h}}}, 4^+_{\tilde h}) 
& = & \left({\kappa \over 2} \right)^{2}  
s_{12} \, A_4(1_g^-,2_g^-,3_g^+,4_g^+) \, 
A_4(1_{\tilde g}^-,2_g^-,4_{\tilde g}^+,3_g^+) \,  \nonumber \\
& = &
\left({\kappa \over 2} \right)^{2} 
s_{12} \times {s_{12}\over s_{23}}  \times \sqrt{s_{12} \over s_{24}}  \,,
\label{GavitinoAmpl}
\end{eqnarray}
where the subscript $\tilde h$ signifies a spin 3/2 gravitino and
$\tilde g$ signifies a spin 1/2 gluino.  As a more subtle example,
the scattering of fundamental representation quarks by gluons via
graviton exchange also has a KLT factorization:
\begin{eqnarray}
{\cal M}_4^{\rm ex}(1_q^{-i_1},&& 
 2_{\bar q}^{+{\bar\imath}_2}, 3_g^{-a_3}, 4_g^{+a_4}) 
\nonumber\\
&&  = 
\Bigl({\kappa \over 2}\Bigr)^2
  s_{12} A_4(1_s^{i_1}, 2_s^{{\bar\imath}_2}, 3_Q^{-a_3}, 4_{\bar Q}^{+a_4}) 
         A_4(1_q^-, 2_{\bar q}^+, 4_{\bar Q}^+, 3_Q^-) \nonumber \\
&& = 
\Bigl({\kappa \over 2}\Bigr)^2  {\sqrt{|s_{13}^3 s_{23}|} \over s_{12}} \;
 \delta_{i_1}{}^{{\bar\imath}_2}\delta^{a_3a_4} \,,
\label{qqggExchange}
\end{eqnarray}
where $q$ and $Q$ are distinct massless fermions. In this equation, the
gluons are factorized into products of fermions. On the
right-hand side the group theory indices $(i_1, \bar\imath_2, a_3,
a_4)$ are interpreted as global flavor indices but on the
left-hand side they should be interpreted as color indices of local
gauge symmetry.  As a check, in Ref.~\cite{Square}, for both
amplitudes (\ref{GavitinoAmpl}) and (\ref{qqggExchange}), ordinary
gravity Feynman rules were used to explicitly verify that the
expressions for the amplitudes are correct.

Cases with multiple fermion pairs are more involved. In particular,
for the KLT factorization to work in general, auxiliary rules for
assigning global charges in the color-ordered amplitudes appear to be
necessary.  This is presumably related to the intricacies associated
with fermions in string theory~\cite{Friedan}.

When an underlying string theory does exist, such as for the case of
maximal supergravity discussed in
section~\ref{section:divergence_properties}, then the KLT equations
necessarily must hold for all amplitudes in the field theory limit.
The above examples, however, demonstrate that the KLT factorization of
amplitudes is not restricted only to the cases where an underlying string
theory exists.

\subsection{Soft and Collinear Properties of Gravity Amplitudes 
from Gauge Theory}
\label{subsection:soft_collinear}

The analytic properties of gravity amplitudes as momenta become either
soft $ (k_i \rightarrow 0)$ or collinear ($k_j$ parallel to $k_j$) are
especially interesting because they supply a simple demonstration of
the tight link between the two theories.  Moreover, these analytic
properties are crucial for constructing and checking gravity
amplitudes with an arbitrary number of external legs.  The properties
as gravitons become soft have been known for a long
time~\cite{WeinbergSoftG,BGK} but the collinear properties were first
obtained using the known collinear properties of gauge theories together
with the KLT relations.

Helicity amplitudes in quantum chromodynamics have a well-known
behavior as momenta of external legs become collinear or
soft~\cite{ManganoReview,Review}.  For the collinear case, at
tree-level in quantum chromodynamics when two nearest neighboring
legs in the color-stripped amplitudes become collinear, {\it e.g.,} $k_1
\rightarrow z P$, $k_2 \rightarrow (1-z) P$, and $P = k_1 + k_2$, 
the amplitude behaves as~\cite{ManganoReview}:
\begin{equation}
A_n^{\rm tree}(1,2,\ldots,n)\ \mathop{\longrightarrow}^{k_1 \parallel k_2}\ 
 \sum_{\lambda = \pm} 
{\mathop{\rm Split}\nolimits}_{-\lambda}^{\rm QCD\ tree}(1,2) \, 
    A_{n-1}^{\rm tree}(P^\lambda,3,\ldots,n)\,.
\label{YMCollinear}
\end{equation}
The function ${\mathop{\rm Split}\nolimits}_{-\lambda}^{\rm QCD\
tree}(1,2)$ is a splitting amplitude, and $\lambda$ is the helicity of
the intermediate state $P$.  (The other helicity labels are implicit.)
The contribution given in Eq.~(\ref{YMCollinear}) is singular for
$k_1$ parallel to $k_2$; other terms in the amplitude are suppressed
by a power of $\sqrt{s_{12}}$, which vanishes in the collinear limit, 
compared to the ones in
Eq.~(\ref{YMCollinear}).  For the pure glue case, one such splitting
amplitude is
\begin{equation}
{\mathop{\rm Split}\nolimits}_-^{\rm QCD\ tree}(1^+,2^+) 
= {1\over \sqrt{z (1-z)}} \, {1\over \langle {1}{2} \rangle} \,,
\label{YMSplitExample}
\end{equation}
where 
\begin{equation}
\langle j l \rangle =  
\sqrt{s_{ij}}\; e^{i \phi_{jl}}\,, \, \hskip 1.5 cm 
[j l]  = -\sqrt{s_{ij}}\; e^{-i \phi_{jl}}\,,
\label{SpinorsDefs}
\end{equation}
are spinor inner products, and $\phi_{jl}$ is a momentum-dependent
phase that may be found in, for example, Ref.~\cite{ManganoReview}.
In general, it is convenient to express splitting amplitudes in terms
of these spinor inner products. The `$+$' and `$-$' labels refer to
the helicity of the outgoing gluons.  Since the spinor inner products
behave as $\sqrt{s_{ij}}$, the splitting amplitudes develop
square-root singularities in the collinear limits.  If the two
collinear legs are not next to each other in the color ordering, then
there is no singular contribution, {\it e.g.} no singularity develops
in $A_n^{\rm tree}(1,2,3,\ldots,n)$ for $k_1$ collinear to $k_3$.

{}From the structure of the KLT relations it is clear that a universal
collinear behavior similar to Eq.~(\ref{YMCollinear}) must hold for 
gravity 
since gravity amplitudes can be obtained from gauge theory ones.  The
KLT relations give a simple way to determine the gravity splitting
amplitudes, ${\mathop{\rm Split}\nolimits}^{\rm gravity\, tree}$.  The
value of the splitting amplitude may be obtained by taking the
collinear limit of two of the legs in, for example, the five-point
amplitude.  Taking $k_1$ parallel to $k_2$ in the five-point relation
(\ref{KLTFivePoint}) and using Eq.~(\ref{GravityCouplingAmplitude}) 
yields:
\begin{equation}
{\cal M}_5^{\rm tree}(1,2,3,4,5)\ 
\mathop{\longrightarrow}^{k_1 \parallel k_2}\ 
{\kappa \over 2}  \sum_{\lambda = \pm} 
{\mathop{\rm Split}\nolimits}_{-\lambda}^{\rm gravity\, tree}(1,2) \, 
   {\cal  M}_{4}^{\rm tree}(P^\lambda,3,4,5)\,,
\label{GravCollinearFive}
\end{equation}
where 
\begin{equation}
{\mathop{\rm Split}\nolimits}^{\rm gravity\ tree}(1,2) = 
-s_{12} \times {\mathop{\rm Split}\nolimits}^{\rm QCD\ tree}(1,2) \times 
{\mathop{\rm Split}\nolimits}^{\rm QCD\ tree}(2,1) \,.
\label{GravTreeColl}
\end{equation}
More explicitly, using Eq.~(\ref{YMSplitExample}) then gives: 
\begin{equation}
{\mathop{\rm Split}\nolimits}_{-}^{\rm gravity\ tree}(1^+,2^+) =
{ - 1\over z (1-z)} {[{1}{2}] \over \langle {1}{2}\rangle }\,.
\label{GravTreeCollExample}
\end{equation}
Using the KLT relations at $n$-points, it is not difficult to 
verify that the splitting behavior is universal for an 
arbitrary number of external legs, {\it i.e.}:
\begin{equation}
{\cal M}_n^{\rm tree}(1,2,\ldots,n)\ 
\mathop{\longrightarrow}^{k_1 \parallel k_2}\ 
{\kappa \over 2}  \sum_{\lambda = \pm} 
{\mathop{\rm Split}\nolimits}_{-\lambda}^{\rm gravity\, tree}(1,2) \, 
   {\cal  M}_{n-1}^{\rm tree}(P^\lambda,3,\ldots,n)\,.
\label{GravCollinear}
\end{equation}
(Since the KLT relations are not manifestly crossing-symmetric, it is
simpler to check this formula for some legs being collinear rather
than others; at the end all possible combinations of legs must give
the same results, though.)  The general structure holds for any
particle content of the theory because of the general applicability of
the KLT relations.

\begin{figure}[h]
  \def\epsfsize#1#2{0.6#1} \centerline{\epsfbox{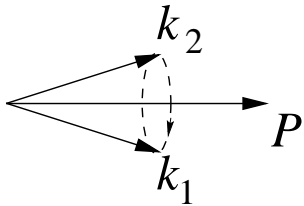}}
  \caption{\it As two momenta become collinear a gravity amplitude
  develops a phase singularity that can be detected by rotating
  the two momenta around the axis formed by their sum.}
  \label{figure:RotCol}
\end{figure}

In contrast to the gauge theory splitting amplitude
(\ref{YMSplitExample}), the gravity splitting amplitude
(\ref{GravTreeCollExample}) is not singular in the collinear limit.
The $s_{12}$ factor in Eq.~(\ref{GravTreeColl}) has canceled the
pole. However, a phase singularity remains from the form of the spinor
inner products given in Eq.~(\ref{SpinorsDefs}), which distinguishes
terms with the splitting amplitude from any others.  In
Eq.~(\ref{SpinorsDefs}), the phase factor $\phi_{12}$ rotates by
$2\pi$ as $\vec k_1$ and $\vec k_2$ rotate once around their sum $\vec
P$ as shown in Fig.~\ref{figure:RotCol}.  The ratio of spinors in
Eq.~(\ref{GravTreeCollExample}) then undergoes a $4\pi$ rotation
accounting for the angular-momentum mismatch of 2$\hbar$ between the
graviton $P^+$ and the pair of gravitons $1^+$ and $2^+$.  In the gauge
theory case, the terms proportional to the splitting
amplitudes~(\ref{YMCollinear}) dominate the collinear limit.
In the gravitational formula ~(\ref{GravCollinear}),
there are other terms of the same magnitude as $[{1}{2}]/
\langle{1}{2}\rangle$ as $s_{12} \to 0$. However, these non-universal
terms do not acquire any additional phase as the collinear vectors
$\vec k_1$ and $\vec k_2$ are rotated around each other. Thus, they
can be separated from the universal terms.  The collinear limit of any
gravity tree amplitude must contain the universal terms given in
Eq.~(\ref{GravCollinear}) thereby putting a severe restriction on the
analytic structure of the amplitudes.

% FIGURE
%%%%%%%%%%%%%%%%%%%%%%%%%%% 
%%%%%%%%%%%%%%%%%%%%%%%%%%

Even for the well-studied case of momenta becoming soft one may again use
the KLT relation to extract the behavior and to rewrite it in terms of
the soft behavior of gauge theory amplitudes.  Gravity tree amplitudes
have the well known behavior~\cite{WeinbergSoftG},
\begin{equation}
{\cal M}_n^{\rm tree}(1,2,\ldots,n^+)\ \mathop{\longrightarrow}^{k_n\to0}\
   {\kappa\over2} \, {\mathop {\cal S} \nolimits}^{\rm gravity}(n^+) \times\ 
   {\cal M}_{n-1}^{\rm tree}(1,2,\ldots,n-1) \,,
\label{GravTreeSoft}
\end{equation}
as the momentum of graviton $n$, becomes soft.  In
Eq.~(\ref{GravTreeSoft}) the soft graviton is taken to carry positive
helicity; parity can be used to obtain the other helicity case. 

One can obtain the explicit form of the soft factors directly from
the KLT relations, but a more symmetric looking soft factor can be
obtained by first expressing the three-graviton vertex in terms of a
Yang-Mills three vertex~\cite{BernGrant}.  (See
Eq.~(\ref{BernGrantVertex}).) This three-vertex can then be used to 
directly construct the soft factor.  The result is a simple formula
expressing the universal function describing soft gravitons in terms of
the universal functions describing soft gluons~\cite{BernGrant}:
\begin{equation}
{\mathop {\cal S} \nolimits}^{\rm gravity}(n^+) = \sum_{i=1}^{n-1} s_{ni} \,
          {\mathop {\cal S} \nolimits}^{\rm QCD}(q_l, n^+, i) \times 
          {\mathop {\cal S} \nolimits}^{\rm QCD}(q_r, n^+, i) \,,
\label{FinalSoftGrav}
\end{equation}
where
\begin{equation}
{\mathop {\cal S} \nolimits}^{\rm QCD}(a, n^+, b) 
= {\langle {a}{b}\rangle \over 
     \langle{a}{n}\rangle \langle{n}{b}\rangle}\,,
\end{equation}
is the eikonal factor for a positive helicity soft gluon in QCD
labeled by $n$, and $a$ and $b$ are labels for legs neighboring the
soft gluon.  In Eq.~(\ref{FinalSoftGrav}) the momenta $q_l$ and $q_r$
are arbitrary null ``reference'' momenta.  Although not manifest, the
soft factor (\ref{FinalSoftGrav}) is independent of the choices of
these reference momenta.  By choosing $q_l = k_1$ and $q_r = k_{n-1}$
the form of the soft graviton factor for $k_n \rightarrow 0$ used in,
for example, Refs.~\cite{BGK,AllPlusGrav,MHVGrav} is recovered.  The
important point is that in the form (\ref{FinalSoftGrav}), the
graviton soft factor is expressed directly in terms of the QCD gluon
soft factor.  Since the soft amplitudes for gravity are expressed in
terms of gauge theory ones, the probability of emitting a soft
graviton can also be expressed in terms of the probability of emitting
a soft gluon.

One interesting feature of the gravitational soft and collinear
functions is that, unlike the gauge theory case, they do not suffer
any quantum corrections~\cite{MHVGrav}.  This is due to the
dimensionful nature of the gravity coupling $\kappa$, which causes any
quantum corrections to be suppressed by powers of a vanishing
kinematic invariant. The dimensions of the coupling constant must be
absorbed by additional powers of the kinematic invariants appearing in
the problem, which all vanish in the collinear or soft limits.  This
observation is helpful because it can be used to put severe
constraints on the analytic structure of gravity amplitudes at any
loop order.

\newpage
%========================================================================

\section{The Einstein-Hilbert Lagrangian and Gauge Theory}
\label{section:EinsteinHilbert}

Consider the Einstein-Hilbert and Yang-Mills Lagrangians,
\begin{equation}
L_{\rm EH} = {2 \over \kappa^2} \sqrt{-g} R \,, \hskip 2 cm 
L_{\rm YM} = - {1\over 4} F^a_{\mu\nu} F^{a\mu\nu}\,, \hskip 1 cm 
\end{equation}
where $R$ is the usual scalar curvature and $F_{\mu\nu}^a$ is the
Yang-Mills field strength. An inspection of these two Lagrangians does
not reveal any obvious factorization property that might explain the
KLT relations.  Indeed, one might be tempted to conclude that the KLT
equations could not possibly hold in pure Einstein gravity.  However, 
although somewhat obscure, the Einstein-Hilbert Lagrangian can in fact be 
rearranged into a form that is  compatible with the KLT relations (as 
argued in this section). Of course,  there should be such a rearrangement, 
given that in the low energy limit pure graviton tree
amplitudes in string theory should match those of Einstein gravity.
All other string states either decouple or cannot enter as
intermediate states in pure graviton amplitudes because of
conservation laws.  Indeed, explicit calculations using ordinary
gravity Feynman rules confirm this to be
true~\cite{Sannan86,BernGrant,Square}. (In loops, any state of the string
that survives in the low energy limit will in fact contribute, but
in this section only tree amplitudes are being considered.)

One of the key properties exhibited by the KLT relations
(\ref{KLTFourPoint}) and (\ref{KLTFivePoint}) is the separation of
graviton space-time indices into `left' and `right' sets. This is a
direct consequence of the factorization properties of closed strings
into open strings.  Consider the graviton field, $h_{\mu\nu}$.  We
define the $\mu$ index be a ``left'' index and the $\nu$ index to be a
``right'' one.  In string theory, the ``left'' space-time indices would
arise from the world-sheet left-mover oscillator and the ``right'' ones
from the right-mover oscillators.  Of course, since $h_{\mu\nu}$ is a
symmetric tensor it does not matter which index is assigned to the
left or to the right.  In the KLT relations each of the two indices of
a graviton are associated with two distinct gauge theories. For
convenience, we similarly call one of the gauge theories the ``left''
one and the other the ``right'' one. Since the indices from each
gauge theory can never contract with the indices of the other gauge
theory, it must be possible to separate all the indices appearing in a
gravity amplitude into left and right classes such that the ones in
the left class only contract with left ones and the ones in the right
class, only with right ones. 

This was first noted by Siegel, who observed that it should be possible
to construct a complete field theory formalism that naturally reflects the
left-right string theory factorization of space-time indices.  In a
set of remarkable papers~\cite{Siegel93A,Siegel93B,Siegel94}, he
constructed exactly such a formalism.  With appropriate gauge choices,
indices separate exactly into ``right'' and ``left'' categories, which do
not contract with each other. This does not provide
a complete explanation of the KLT relations, since one would still need
to demonstrate that the gravity amplitudes can be expressed directly
in terms of gauge theory ones.  Nevertheless, this formalism is
clearly a sensible starting point for trying to derive the KLT
relations directly from Einstein gravity.  Hopefully, this will be the
subject of future studies, since it may lead to a deeper understanding
of the relationship of gravity to gauge theory.  A Lagrangian with the
desired properties could, for example, lead to more general relations
between gravity and gauge theory classical solutions.

Here we outline a more straightforward order-by-order rearrangement of
the Einstein-Hilbert Lagrangian, making it compatible with the KLT
relations~\cite{BernGrant}.  A useful side-benefit is that this
provides a direct verification of the KLT relations up to
five points starting from the Einstein-Hilbert Lagrangian in its usual
form.  This is a rather non-trivial direct verification of the KLT
relations, given the algebraic complexity of the gravity Feynman
rules.

In conventional gauges, the difficulty of factorizing the
Einstein-Hilbert Lagrangian into left and right parts is already
apparent in the kinetic terms. In De~Donder gauge, for example, the
quadratic part of the Lagrangian is
\begin{equation}
L_2 =  - \frac{1}{2} h_{\mu\nu} \partial^2 h^{\mu\nu} 
+ \frac{1}{4} h_{\mu}{}^{\mu} \partial^2 h_{\nu}{}^{\nu} \,,
\label{QuadraticLagrang}
\end{equation}
so that the propagator is the one given in
Eq.~(\ref{GravityPropagator}).  Although the first term is acceptable
since left and right indices do not contract into each other, the
appearance of the trace $h_{\mu}{}^{\mu}$ in
Eq.~(\ref{QuadraticLagrang}) is problematic since it contracts a left
graviton index with a right one. (The indices are raised and lowered
using the flat space metric $\eta_{\mu\nu}$ and its inverse.)

In order for the kinematic term~(\ref{QuadraticLagrang}) to be
consistent with the KLT equations, all terms which contract a ``left''
space-time index with a ``right'' one need to be eliminated.  A useful
trick for doing so is to introduce a ``dilaton'' scalar field that can be used to
remove the graviton trace from the quadratic terms in the Lagrangian.
The appearance of the dilaton as an auxiliary field to help rearrange
the Lagrangian is motivated by string theory, which requires the
presence of such a field.  Following the discussion of
Refs.~\cite{BDS,BernGrant}, consider instead a Lagrangian for gravity
coupled to a scalar:
\begin{equation}
L_{\rm EH} =   {2\over \kappa^2} \sqrt{-g} \, 
       R +  \sqrt{-g} \, \partial^\mu\phi\partial_\mu\phi \,.
\end{equation}
Since the auxiliary field $\phi$ is quadratic in the Lagrangian, it
does not appear in any tree diagrams involving only external
gravitons~\cite{BernGrant}.  It therefore does not alter the tree
$S$-matrix of purely external gravitons.  (For theories containing
dilatons one can allow the dilaton to be an external physical
state.)  In De~Donder gauge, for example, taking $g_{\mu\nu} =
\eta_{\mu\nu} + \kappa h_{\mu\nu}$, the quadratic part of the
Lagrangian including the dilaton is:
\begin{equation}
L_2 = - \frac{1}{2} h_{\mu\nu} \partial^2 h^{\mu\nu} 
+ \frac{1}{4} h_{\mu}{}^{\mu} \partial^2 h_{\nu}{}^{\nu}
-  \phi \partial^2 \phi \,.
\end{equation}
The term involving $h_{\mu}{}^{\mu}$ can be eliminated with the
field redefinitions,
\begin{equation}
 h_{\mu\nu} \rightarrow h_{\mu\nu} 
        + \eta_{\mu\nu}{\sqrt{\frac{2}{D-2}}}\, \phi\,,
\end{equation}
and
\begin{equation}
 \phi \rightarrow \frac{1}{2} h_{\mu}{}^{\mu} + \sqrt{\frac{D-2}{2}}\, \phi\,,
\label{FieldRedefA}
\end{equation}
yielding
\begin{equation}
L_2 \rightarrow 
- \frac{1}{2} h^{\mu}{}_{\nu} \partial^2 h_{\mu}{}^{\nu} 
+  \phi \partial^2 \phi\,.
\end{equation}
One might be concerned that the field redefinition might alter gravity
scattering amplitudes.  However, because this field redefinition does
not alter the trace-free part of the graviton field it cannot change
the scattering amplitudes of traceless
gravitons~\cite{BernGrant}. 

Of course, the rearrangement of the quadratic terms is only the first
step.  In order to make the Einstein-Hilbert Lagrangian consistent
with the KLT factorization, a set of field variables should exist where
all space-time indices can be separated into ``left'' and ``right''
classes. To do so, all terms of the form
\begin{equation}
h_{\mu}{}^{\mu}\, , \hskip .5 cm 
h_{\mu}{}^{\nu} h_{\nu}{}^{\lambda} h_{\lambda}{}^{\mu} \,, \hskip .5 cm 
\cdots, \hskip .5 cm
\label{BadTerms}
\end{equation}
need to be eliminated since they contract left indices with right
 ones.  A field redefinition that accomplishes this
 is~\cite{BernGrant}:
\begin{equation}
g_{\mu\nu}  = e^{\sqrt{\frac{2}{D-2}} \kappa \phi} e^{\kappa\, h_{\mu\nu}} 
 \equiv 
 e^{\sqrt{\frac{2}{D-2}} \,\kappa \phi} 
\Bigl(\eta_{\mu\nu} + \kappa h_{\mu\nu} + {\textstyle \kappa^2\over 2} 
                    h_{\mu}{}^{\rho} h_{\rho\nu} + \cdots \Bigr) \,. 
\end{equation}
This field redefinition was explicitly checked in
Ref.~\cite{BernGrant} through ${\cal O}(h^6)$, to eliminate all terms
of the type in Eq.~(\ref{BadTerms}), before gauge fixing.  However,
currently there is no formal understanding of why this field variable
choice eliminates terms that necessarily contract left and right
indices.

It turns out that one can do better by performing further field
redefinitions and choosing a particular non-linear gauge.  The
explicit forms of these are a bit complicated and may be found in
Ref.~\cite{BernGrant}.  With a particular gauge choice it is possible
to express the off-shell three-graviton vertex in terms of Yang-Mills
three vertices:
\begin{eqnarray}
i G^{\mu_1 \nu_1, \mu_2\nu_2, \mu_3\nu_3}(k_1, k_2 , k_3)
& = &- {i \over 2} 
\Bigl( {\kappa \over 2} \Bigr)
\Bigl[
V_{\rm GN}^{\mu_1 \mu_2 \mu_3}(k_1, k_2, k_3) 
\times V_{\rm GN}^{\nu_1 \nu_2 \nu_3}(k_1, k_2, k_3) \nonumber \\
&&  \null
+ V_{\rm GN}^{\mu_2 \mu_1 \mu_3}(k_2, k_1, k_3) \times 
  V_{\rm GN}^{\mu_2 \mu_1 \mu_3}(k_2, k_1, k_3) \Bigr]\,, 
\label{BernGrantVertex}
\end{eqnarray}
where
\begin{equation}
 V^{\mu\nu\rho}_{\rm GN}(k_1,k_2,k_3) = 
       i\sqrt{2}  \bigl( k_1^\rho \, \eta^{\mu\nu}
            + k_2^\mu \eta^{\nu\rho} + k_3^\nu \eta^{\rho\mu} \bigr)\,,
\end{equation}
is the color-ordered Gervais-Neveu~\cite{GN} gauge Yang-Mills three-
vertex, from which the color factor has been stripped.  This is not
the only possible reorganization of the three-vertex that respects
the KLT factorization.  It just happens to be a particularly simple
form of the vertex.  For example, another gauge that has a three-vertex 
that factorizes into products of color-stripped Yang-Mills
three-vertices is the
background-field~\cite{tHooft75,DeWitt81,Background} version of
De~Donder gauge for gravity and Feynman gauge for QCD.  (However,
background field gauges are meant for loop effective actions and not
for tree-level $S$-matrix elements.)  Interestingly, these gauge
choices have a close connection to string theory~\cite{GN,Mapping}.

The above ideas represent some initial steps in reorganizing the
Einstein-Hilbert Lagrangian so that it respects the KLT relations.  An
important missing ingredient is a derivation of the KLT equations
starting from the Einstein-Hilbert Lagrangian (and also when matter
fields are present).

\newpage
%======================================================================

\section{From Trees to Loops}
\label{section:trees_to_loops}

In this section, the above discussion is extended to quantum loops
through use of $D$-dimensional
unitarity~\cite{Bern94SusyFour,Bern95SusyFour,BernMorgan,Review,Rozowsky}.
The KLT relations provide gravity amplitudes only at tree level;
$D$-dimensional unitarity then provides a means of obtaining quantum
loop amplitudes. In perturbation theory this is tantamount to
quantizing the theory since the complete scattering matrix can, at least in
principle, be systematically constructed this way.  Amusingly, this
bypasses the usual formal
apparatus~\cite{Faddeev74,Fradkin75,Batalin77,Henneaux85} associated
with quantizing constrained systems.  More generally, the unitarity
method provides a way to systematically obtain the complete set of
quantum loop corrections order-by-order in the perturbative expansion
whenever the full analytic behavior of tree amplitudes as a function
of $D$ is known.  It always works when the particles in the theory are
all massless.  The method is well tested in explicit calculations and
has, for example, recently been applied to state-of-the-art perturbative QCD
loop computations~\cite{Bern00QCDApplications,Bern02QCDApplications}.

\begin{figure}[h]
  \def\epsfsize#1#2{0.9#1} \centerline{\epsfbox{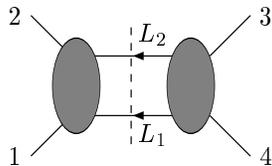}}
  \caption{\it The two-particle cut at one loop in the channel
  carrying momentum $k_1+k_2$. The blobs represent tree
  amplitudes.}  
\label{figure:TwoParticle}
\end{figure}

In quantum field theory the $S$-matrix links initial and final states.
A basic physical property is that the $S$ matrix must be
unitary~\cite{Mandelstam58,Landau59,Mandelstam59,Cutkosky60}: $S^\dag
S = 1$.  In perturbation theory the Feynman diagrams describe a
transition matrix $T$ defined by $S \equiv 1 + i T$, so that the
unitarity condition reads
\begin{equation}
 2 \, {\rm Im}\, T_{if}\ =\ \sum_j T^*_{ij} T_{jf} \, , 
\label{BasicUnitarity}
\end{equation}
where $i$ and $f$ are initial and final states, and the ``sum'' is over
intermediate states $j$ (and includes an integral over intermediate
on-mass-shell momenta).  Perturbative unitarity consists of expanding
both sides of Eq.~(\ref{BasicUnitarity}) in terms of coupling
constants, $g$ for gauge theory and $\kappa$ for gravity, and
collecting terms of the same order.  For example, the imaginary (or
absorptive) parts of one-loop four-point amplitudes, which is order
$\kappa^4$ in gravity, are given in terms of the product of two
four-point tree amplitudes, each carrying a power of $\kappa^2$.  This
is then summed over all two-particle states that can appear and
integrated over the intermediate phase space. (See
Fig.~\ref{figure:TwoParticle}.)

% FIGURE
%%%%%%%%%%%%%%%%%%%%%%%%%%% 
%%%%%%%%%%%

This provides a means of obtaining loop amplitudes from tree
amplitudes.  However, if one were to directly apply
Eq.~(\ref{BasicUnitarity}) in integer dimensions one would encounter a
difficulty with fully reconstructing the loop scattering amplitudes.
Since Eq.~(\ref{BasicUnitarity}) gives only the imaginary part one
then needs to reconstruct the real part.  This is traditionally done
via dispersion relations, which are based on the analytic properties of
the $S$ matrix~\cite{Mandelstam58,Landau59,Mandelstam59,Cutkosky60}.
However, the dispersion integrals do not generally converge. This
leads to a set of subtraction ambiguities in the real part.  These
ambiguities are related to the appearance of rational functions with
vanishing imaginary parts, $R(s_{i})$, where the $s_{i}$ are the
kinematic variables for the amplitude.

A convenient way to deal with this
problem~\cite{Bern94SusyFour,Bern95SusyFour,BernMorgan,Review,Rozowsky}
is to consider unitarity in the context of dimensional
regularization~\cite{HV,vanNeerven}.  By considering the amplitudes as
an analytic function of dimension, at least for a massless theory,
these ambiguities are not present, and the only remaining ambiguities
are the usual ones associated with renormalization in quantum field
theory.  The reason there can be no ambiguity relative to Feynman
diagrams follows from simple dimensional analysis for amplitudes in
dimension $D=4-2 \epsilon$.  With dimensional regularization,
amplitudes for massless particles necessarily acquire a factor of
$(-s_{i})^{-\epsilon}$ for each loop, from the measure $\int d^DL$.
For small $\epsilon$, $(-s_{i})^{-\epsilon} \, R(s_{i}) = R(s_{i}) -
\epsilon \, \ln (-s_{i}) \, R(s_{i}) + {\cal O}(\epsilon^2)$, so every
term has an imaginary part (for some $s_{i}>0$), though not
necessarily in terms which survive as $\epsilon\rightarrow 0$.  Thus,
the unitarity cuts evaluated to ${\cal O}(\epsilon)$ provide
sufficient information for the complete reconstruction of an
amplitude.  Furthermore, by adjusting the specific rules for the
analytic continuation of the tree amplitudes to $D$-dimensions one can
obtain results in the different varieties of dimensional
regularization, such as the conventional one~\cite{CollinsBook}, the
t'~Hooft-Veltman scheme~\cite{HV}, dimensional
reduction~\cite{Siegel79DR}, and the four-dimensional helicity
scheme~\cite{Long,TwoloopFDH}.

It is useful to view the unitarity-based technique as an alternate way
of evaluating sets of ordinary Feynman diagrams by collecting together
gauge-invariant sets of terms containing residues of poles in the
integrands corresponding to those of the propagators of the cut
lines. This gives a region of loop-momentum integration where the cut
loop momenta go on shell and the corresponding internal lines become
intermediate states in a unitarity relation.  From this point of view,
even more restricted regions of loop momentum integration may be
considered, where additional internal lines go on mass shell.  This
amounts to imposing cut conditions on additional internal lines.  In
constructing the full amplitude from the cuts it is convenient to use
unrestricted integrations over loop momenta, instead of phase space
integrals, because in this way one can obtain simultaneously both the
real and imaginary parts. The generalized cuts then allow one to
obtain multi-loop amplitudes directly from combinations of tree
amplitudes.

As a first example, the generalized cut for a one-loop four-point
amplitude in the channel carrying momentum $k_1 + k_2$, as shown in
Fig.~\ref{figure:TwoParticle}, is given by: 
\begin{equation}
 \sum_{\rm states} \int {d^D L_1 \over (4 \pi)^D} \, 
  {i\over L_1^2} \,
{\cal A}_{4}^{\rm tree}(-L_1,1,2,L_2) \,{i \over L_2^2}\,
{\cal A}_{4}^{\rm tree} (-L_2,3,4,L_1) \Bigr|_{\rm2\ cut} \,, 
\label{BasicCutEquation}
\end{equation}
where $L_2 = L_1 - k_1 - k_2$, and the sum runs over all physical
states of the theory crossing the cut.  In this generalized cut, the
on-shell conditions $L_1^2 = L_2^2 = 0$ are applied even though the
loop momentum is unrestricted.  In addition, any physical state
conditions on the intermediate particles should also be included.  The
real and imaginary parts of the integral functions that do have cuts
in this channel are reliably computed in this way. However, the use of the
on-shell conditions inside the unrestricted loop momentum integrals does
introduce an arbitrariness in functions that do not have cuts in this
channel. Such integral functions should instead be obtained from cuts
in the other two channels.

% FIGURE
%%%%%%%%%%%%%%%%%%%%%%%%%%% 
\begin{figure}[h]
  \def\epsfsize#1#2{0.9#1} \centerline{\epsfbox{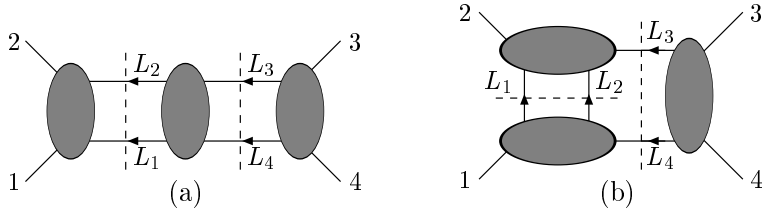}}
  \caption{\it Two examples of generalized cuts. Double two-particle
  cuts of a two-loop amplitude are shown. This separates the amplitude into a
  product of three tree amplitudes, integrated over loop momenta.  The
  dashed lines represent the cuts. }
\label{figure:DoubleDoubleAmpl}
\end{figure}
%%%%%%%%%%%

A less trivial two-loop example of a generalized ``double'' two particle
cut is illustrated in Fig.~\ref{figure:DoubleDoubleAmpl}(a).
The product of tree amplitudes appearing in this cut is:
\begin{eqnarray}
&&
\sum_{\rm states} {\cal A}_4^{\rm tree}(1,2, -L_2, -L_1)  \nonumber \\
&& \hskip 1.8 cm 
\null \times {\cal A}_4^{\rm tree}(L_1, L_2, -L_3, -L_4)
\null \times {\cal A}_4^{\rm tree}(L_4, L_3, 3, 4) 
 \Bigr|_{\rm 2\times2\hbox{\small -} cut} \,, 
\label{DoubleDoubleCut}
\end{eqnarray}
where the loop integrals and cut propagators have been suppressed for
convenience.  In this expression the on-shell conditions
$L_i^2 = 0$ are imposed on the $L_i$, $i=1,2,3,4$ appearing on the
right-hand side. This double cut
may seem a bit odd from the traditional viewpoint in which each 
cut can be interpreted as the imaginary part of the integral.
It should instead be understood as a means to obtain part
of the information on the structure of the integrand of the two-loop
amplitude.  Namely, it contains the information on all integral
functions where the cut propagators are not cancelled. 
 There are, of
course, other generalized cuts at two loops.  For example, in
Fig.~\ref{figure:DoubleDoubleAmpl}(b) a different arrangement of the
cut trees is shown.

Complete amplitudes are found by combining the various cuts into a
single function with the correct cuts in all channels.  This method
works for any theory where the particles can be taken to be
massless and where the tree amplitudes are known as an analytic
function of dimension.  The restriction to massless amplitudes is
irrelevant for the application of studying the ultra-violet
divergences of gravity theories.  In any case, gravitons and their
associated superpartners in a supersymmetric theory are massless.
(For the case with masses present the extra technical complication has
to do with the appearance of functions such as $m^{-2\epsilon}$ which
have no cuts in any channel. See Ref.~\cite{BernMorgan} for a
description and partial solution of this problem.)  This method has been
extensively applied to the case of one- and two-loop gauge theory
amplitudes~\cite{Bern94SusyFour,Bern95SusyFour,Review,Bern00QCDApplications,
Bern02QCDApplications} and has been carefully cross-checked with Feynman
diagram calculations.  Here, the method is used to obtain loop
amplitudes directly from the gravity tree amplitudes given by the KLT
equations.  In the next section an example of how the method works in
practice for the case of gravity is provided.

\newpage
%=========================================================================

\section{Gravity Loop Amplitudes from Gauge Theory}
\label{section:gravity_loops}

The unitarity method provides a natural means for applying the KLT
formula to obtain loop amplitudes in quantum gravity, since the only
required inputs are tree-level amplitudes valid for $D$-dimensions;
this is precisely what the KLT relations provide.

Although Einstein gravity is almost certainly not a fundamental theory,
there is no difficulty in using it as an effective field
theory~\cite{Weinberg79,Gasser85,Donoghue94,Kaplan95,Manohar96}, in
order to calculate quantum loop corrections.  The particular examples
discussed in this section are completely finite and therefore do not
depend on a cutoff or on unknown coefficients of higher curvature
terms in the low energy effective action.  They are therefore a
definite low energy prediction of {\it any} fundamental theory of
gravity whose low energy limit is Einstein gravity.  (Although they
are definite predictions, there is, of course, no practical means to
experimentally verify them.)  The issue of divergences is deferred to
section~\ref{section:divergence_properties}.

\subsection{One-loop Four-point Example}

As a simple example of how the unitarity method gives loop amplitudes,
consider the one-loop amplitude with four identical helicity gravitons
and a scalar in the loop~\cite{AllPlusGrav,MHVGrav}. The product of
tree amplitudes appearing in the $s_{12}$ channel unitarity cut
depicted in Fig.~\ref{figure:TwoParticle} is:
\begin{equation}
 M_4^{\rm tree}(-L_1^s, 1^+_h, 2^+_h, L_3^s) \times 
M_4^{\rm tree}(-L_3^s, 3^+_h, 4^+_h, L_1^s) \,, 
\end{equation}
where the superscript $s$ indicates that the cut lines are scalars.
The $h$ subscripts on legs $1\ldots 4$ indicate that these are gravitons,
while the ``$+$'' superscripts indicate that they are of plus helicity.
{}From the KLT expressions (\ref{KLTFourPoint}) the gravity tree
amplitudes appearing in the cuts may be replaced with products of
gauge theory amplitudes.  The required gauge theory tree amplitudes,
with two external scalar legs and two gluons, may be obtained
using color-ordered Feynman diagrams and are:
\begin{equation}
A_4^{\rm tree}(-L_1^s,1^+_g,2^+_g,L_3^s) =  i
{\mu^2 \over ( L_1 -k_1)^2} \,,
\end{equation}
\begin{equation}
 A_4^{\rm tree}(-L_1^s, 1^+_g, L_3^s, 2^+_g) 
 = - i {\mu^2} 
 \biggl[{1\over (L_1 -k_1)^2 }
 + {1\over (L_1 -k_2)^2}\biggr] \,.
\end{equation}
The external gluon momenta are four dimensional, but the scalar
momenta $L_1$ and $L_3$ are $D$ dimensional since they will form the
loop momenta. In general, loop momenta will have a non-vanishing
$(-2\epsilon)$-dimensional component $\vec{\mu}$, with
$\vec{\mu}\cdot\vec{\mu} = \mu^2 > 0$.  The factors of $\mu^2$
appearing in the numerators of these tree amplitudes causes them to
vanish as the scalar momenta are taken to be four dimensional, though
they are non-vanishing away from four dimensions.  For simplicity,
overall phases have been removed from the amplitudes.  After inserting
these gauge theory amplitudes in the KLT relation
(\ref{KLTFourPoint}), one of the propagators cancels, leaving
\begin{equation}
 M_4^{\rm tree}(-L_1^s, 1^+_h, 2^+_h, L_3^s) = 
- i \mu^4 
 \biggl[{1\over (L_1 -k_1)^2}
 + {1\over (L_1 -k_2)^2}  \biggr] \,.
\end{equation}
For this cut, one then obtains a sum of box integrals that 
can be expressed as:
\begin{eqnarray}
&& \int {d^D L_1 \over (2 \pi)^D} \, \mu^8 \, {i \over L_1^2} 
 \biggl[{i \over (L_1 -k_1)^2}
 + {i \over (L_1 -k_2)^2}  \biggr] \nonumber \\
&& \hskip 2 cm \times {i \over (L_1 - k_1 - k_2)^2}
 \biggl[{i \over (L_1 + k_3)^2}
 + {i \over (L_1 + k_4)^2}  \biggr] \,.
\end{eqnarray}
By symmetry, since the helicities of all the external gravitons are
identical, the other two cuts also give the same combinations of
box integrals, but with the legs permuted.

\begin{figure}[h]
  \def\epsfsize#1#2{0.5#1}
  \centerline{\epsfbox{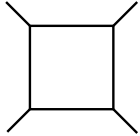}}
  \caption{\it The one-loop box integral. Each internal 
  line in the box corresponds to one of the four Feynman propagators in 
   Eq.~(\ref{OneLoopIntegral}).}
  \label{figure:OneloopIntegral}
\end{figure}

The three cuts can then be combined into a single function that has the correct
cuts in all channels yielding 
\begin{eqnarray}
M_4^{\rm 1 \,loop}(1^+_h, 2^+_h, 3^+_h, 4^+_h)  & = &
 2 \Bigl({\cal I}_4^{\rm 1\, loop}[\mu^8](s_{12},s_{23}) + 
       {\cal I}_4^{\rm 1\, loop}[\mu^8](s_{12},s_{13}) \nonumber \\
& &
\null   + {\cal I}_4^{\rm 1\, loop}[\mu^8](s_{23},s_{13}) \Bigr)\,, 
\label{FourGravAllPlus}
\end{eqnarray}
and where
\begin{equation}
{\cal I}_4^{\rm 1\, loop}[{\cal P}](s_{12},s_{23}) = 
\int {d^D L \over (2\pi)^D}
\, {{\cal P} \over L^2 (L - k_1)^2
       (L - k_1 - k_2)^2
        (L + k_4)^2  } \,,
\label{OneLoopIntegral} 
\end{equation}
is the box integral depicted in
Fig.~\ref{figure:OneloopIntegral} with the external legs arranged in
the order 1234. In Eq.~(\ref{FourGravAllPlus}) ${\cal P}$ is $\mu^8$.
The two other integrals that appear correspond to the two other
distinct orderings of the four external legs.  The overall factor of 2
in Eq.~(\ref{FourGravAllPlus}) is a combinatoric factor due to taking
the scalars to be complex with two physical states.

Since the factor of $\mu^8$ is of ${\cal O}(\epsilon)$, the only non-vanishing
contributions come where the $\epsilon$ from the $\mu^8$ interferes with 
a divergence in the loop integral.  These divergent contributions
are relatively simple  to obtain.  After extracting this contribution from
the integral, the final $D=4$ result for a complex scalar loop, after
reinserting the gravitational coupling, is
\begin{equation}
{\cal M}_4^{\rm 1\, loop} (1^+_h, 2^+_h, 3^+_h, 4^+_h) 
   =  \, {1 \over (4 \pi)^2} \Bigl({\kappa\over 2} \Bigr)^4\, 
     {s_{12}^2 + s_{23}^2 + s_{13}^2 \over 120} \,, 
\label{FourGravPlus}
\end{equation}
in agreement with a calculation done by a different
method relying directly on string theory~\cite{DunbarNorridge95}. 
(As for the previous expressions, the overall phase has been suppressed.)

%FIGURE
%%%%%%%%%%%%%%%%%%%%%%%%%%% 
%%%%%%%%%%%%%%%%%%%%%%%%%%

This result generalizes very simply to the case of any particles in
the loop.  For any theory of gravity, with an arbitrary matter content
one finds:
\begin{equation}
{\cal M}_4^{\rm 1\, loop} (1^+_h, 2^+_h, 3^+_h, 4^+_h)  =  \, {N_s\over 2}
 {1 \over (4 \pi)^2} 
\Bigl({\kappa\over 2} \Bigr)^4\, 
{s_{12}^2 + s_{23}^2 + s_{13}^2 \over 120} \,, 
\label{FourGravPlusGen}
\end{equation}
where $N_s$ is the number of physical bosonic states circulating in
the loop minus the number of fermionic states.  The simplest way to
demonstrate this is by making use of supersymmetry Ward
identities~\cite{Grisaru77SWIB,Parke85SWI,Review}, which provide a set
of simple linear relations between the various contributions showing that
they must be proportional to each other.

\subsection{Arbitrary Numbers of Legs at One Loop}

Surprisingly, the above four-point results can be extended to an
arbitrary number of external legs.  Using the unitarity methods, the
five- and six-point amplitudes with all identical helicity have also
been obtained by direct calculation~\cite{AllPlusGrav,MHVGrav}.  Then
by demanding that the amplitudes have the properties described in
section~\ref{subsection:soft_collinear} for momenta becoming either
soft~\cite{WeinbergSoftG,BGK} or collinear~\cite{AllPlusGrav}, an
ansatz for the one-loop maximally helicity-violating amplitudes for an
arbitrary number of external legs has also been obtained.  These amplitudes
were constructed from a set of building blocks called
``half-soft-function,'' which have ``half'' of the proper behavior as
gravitons become soft.  The details of this construction and the
explicit forms of the amplitudes may be found in
Refs.~\cite{AllPlusGrav,MHVGrav}.

The all-plus helicity amplitudes turn out to be very closely related
to the infinite sequence of one-loop maximally helicity-violating
amplitudes in $N=8$ supergravity.  The two sequences are related by a
curious ``dimension shifting formula.''  In Ref.~\cite{MHVGrav}, a known
dimension shifting formula~\cite{Bern97DimShift} between identical helicity
QCD and $N=4$ super-Yang-Mills amplitudes was used to obtain the
four-, five-, and six-point $N=8$ amplitudes from the identical 
helicity gravity amplitudes using the KLT relations in the 
unitarity cuts.  Armed with these explicit results, the soft and
collinear properties were then used to obtain an ansatz valid for an
arbitrary number of external legs~\cite{MHVGrav}.  This provides a
rather non-trivial illustration of how the KLT relations can be used
to identify properties of gravity amplitudes using known properties of
gauge theory amplitudes.

Interestingly, the all-plus helicity amplitudes are also connected to
self-dual gravity~\cite{Plebanski75,Duff79,Plebanski96} and self-dual
Yang-Mills~\cite{Yang77,DuffIsham80,Lenzov87,Lenzov88,Bardeen96,
Cangemi,ChalmersSiegel}, {\it i.e.} gravity and gauge theory
restricted to self-dual configurations of the respective field
strengths, $R_{\mu\nu\rho\sigma} =
{i\over2}\epsilon_{\mu\nu}{}^{\alpha\beta} R_{\alpha\beta \rho\sigma}$
and $F_{\mu\nu} = {i\over 2} \epsilon_{\mu\nu}{}^{\alpha\beta}
F_{\alpha\beta}$, with $\epsilon_{0123} = +1$.  This connection is
simple to see at the linearized (free field theory) level since a
superposition of plane waves of identical helicity satisfies the
self-duality condition.  The self-dual currents and amplitudes have
been studied at tree and one-loop
levels~\cite{DuffIsham80,Bardeen96,Cangemi,ChalmersSiegel}.  In
particular, Chalmers and Siegel~\cite{ChalmersSiegel} have presented
self-dual actions for gauge theory (and gravity), which reproduce the
all-plus helicity scattering amplitudes at both tree and one-loop
levels.

The ability to obtain exact expressions for gravity loop amplitudes
demonstrates the utility of this approach for investigating quantum
properties of gravity theories.  The next section describes how this
can be used to study high energy divergence properties 
in quantum gravity.

\newpage 
%=============================================================================
\section{Divergence Properties of Maximal Supergravity}
\label{section:divergence_properties}

In general, the larger the number of supersymmetries, the tamer the
ultraviolet divergences because of the tendency for these to cancel
between bosons and fermions in a supersymmetric theory.  In
four-dimensions maximal $N=8$ supergravity may therefore be expected
to be the least divergent of all possible supergravity
theories. Moreover, the maximally supersymmetric gauge theory, $N=4$
super-Yang-Mills, is completely finite~\cite{Stelle81,Mandelstam83,Howe89},
leading one to suspect that the superb ultraviolet properties of $N=4$
super-Yang-Mills would then feed into improved ultra-violet properties
for $N=8$ supergravity via its relation to gauge theory.  This makes
the ultraviolet properties of $N=8$ supergravity the ideal case to
investigate first via the perturbative relationship to gauge theory.

\subsection{One-loop Cut Construction}
\label{subsection:supergravity_cut_construction}

The maximal $N=8$ supergravity amplitudes can be obtained by applying
the KLT equations to express them in terms of maximally supersymmetric
$N=4$ gauge theory amplitudes.  For $N=8$ supergravity, each of the states
of the multiplet factorizes into a tensor product of $N=4$
super-Yang-Mills states, as illustrated in
Eq.~(\ref{StateFactorization}). Applying the KLT equation
(\ref{KLTFourPoint}) to the product of tree amplitudes appearing in
the $s_{12}$ channel two-particle cuts yields:
\begin{eqnarray}
\sum_{N=8 \atop \rm\ states}  
&& M_4^{\rm tree}(-L_1,  1, 2, L_3) \times
  M_4^{\rm tree}(-L_3, 3, 4, L_1) \nonumber \\
&& \hskip 1.5 cm  = 
- s_{12}^2 \!\sum_{N=4 \atop \rm\ states} \!
 A_4^{\rm tree}(-L_1,  1, 2, L_3) \times
     A_4^{\rm tree}(-L_3,  3,4 , L_1) \nonumber \\
\null & & \hskip 1.9 truecm
\times
\!\sum_{N=4 \atop \rm\ states} \!
 A_4^{\rm tree}(L_3, 1, 2 , -L_1) \times
                A_4^{\rm tree}(L_1, 3, 4, -L_3) \,,
\label{GravitySewingStart}
\end{eqnarray}
where the sum on the left-hand side runs over all 256 states in the
$N=8$ supergravity multiplet. On the right-hand side the two sums run
over the 16 states (ignoring color degrees of freedom) of the $N=4$
super-Yang-Mills multiplet: a gluon, four Weyl fermions and six real
scalars.

The $N=4$ super-Yang-Mills tree amplitudes turn out to have a 
particularly simple sewing formula~\cite{BRY},
\begin{eqnarray}
&& \sum_{N=4\atop \rm  states}
 A_4^{\rm tree}(-L_1, 1, 2, L_3) \times
  A_4^{\rm tree}(-L_3, 3, 4, L_1) \nonumber \\
&& \hskip 2 cm 
 =  - i s_{12} s_{23} \, A_4^{\rm tree}(1, 2, 3, 4) \, 
   {1\over (L_1 - k_1)^2 } \, 
   {1\over (L_3 - k_3)^2 } \,,
\label{YangMillsSewing}
\end{eqnarray}
which holds in {\it any} dimension (though some care is required to maintain
the total number of physical states at their four-dimensional values so
as to preserve the supersymmetric cancellations).
The simplicity of this result is due to the high degree of supersymmetry.

Using the gauge theory result~(\ref{YangMillsSewing}), it is a simple
matter to evaluate Eq.~(\ref{GravitySewingStart}). This yields:
\begin{eqnarray}
&& \sum_{N=8 \atop \rm\ states}  M_4^{\rm tree}(-L_1, 1, 2, L_3) \times
  M_4^{\rm tree}(-L_3, 3, 4, L_1) \nonumber \\
&& \hskip 2 cm 
= i s_{12} s_{23} s_{13} M_4^{\rm tree}(1, 2, 3, 4)
 \biggl[{1\over (L_1 - k_1)^2 } + {1\over (L_1 - k_2)^2} \biggr] \nonumber \\
&& \hskip 5.03 cm \times
\biggl[{1\over (L_3 - k_3)^2 } + {1\over (L_3 - k_4)^2} \biggr]\,.
\label{BasicGravCutting}
\end{eqnarray}
The sewing equations for the $s_{23}$ and $s_{13}$ kinematic channels are
similar to that of the $s_{12}$ channel.

Applying Eq.~(\ref{BasicGravCutting}) at one loop to each of the three
kinematic channels yields the one-loop four graviton amplitude of $N=8$
supergravity,
\begin{eqnarray}
 {\cal M}_4^{{\rm 1\, loop}}(1, 2, 3, 4)
& = &  -i \Bigl( {\kappa \over 2}\Bigr)^4 
 s_{12} s_{23} s_{13}\,  M_4^{\rm tree}(1,2,3,4)  
 \Bigl(  {\cal I}_4^{{\rm 1\, loop}}(s_{12},s_{23}) \nonumber\\
&& \hskip 1.3 cm  
       \null    + {\cal I}_4^{{\rm 1\, loop}}(s_{12},s_{13})  
           + {\cal I}_4^{{\rm 1\, loop}}(s_{23},s_{13})  \Bigr) \,,
\end{eqnarray}
in agreement with previous results~\cite{GSB}.  The gravitational
 coupling $\kappa$ has been reinserted into this expression.  The
 scalar integrals are defined in Eq.~(\ref{OneLoopIntegral}),
 inserting ${\cal P} = 1$.  This is a standard integral appearing in
 massless field theories; the explicit value of this integral may be
 found in many articles, including Refs.~\cite{GSB,Long}.  This result
 actually holds for any of the states of $N=8$ supergravity, not just
 external gravitons.  It is also completely equivalent to the
 result one obtains with covariant Feynman diagrams including
 Fadeev-Popov~\cite{Faddeev} ghosts and using regularization by
 dimensional reduction~\cite{Siegel79DR}.  The simplicity of this
 result is due to the high degree of supersymmetry.  A generic
 one-loop four-point gravity amplitude can have up to eight powers of
 loop momenta in the numerator of the integrand; the supersymmetry
 cancellations have reduced it to no powers.

\subsection{Higher Loops}

\begin{figure}[h]
  \def\epsfsize#1#2{0.5#1}
  \centerline{\epsfbox{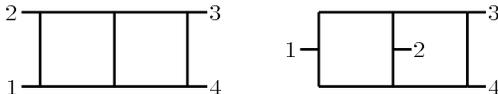}}
  \caption{\it The planar and non-planar scalar integrals, 
   appearing in the two-loop N=8 amplitudes.
   Each internal line represents a scalar propagator.}
  \label{figure:PlanarNonPlanar}
\end{figure}

At two loops, the two-particle cuts are obtained easily by
iterating the one-loop calculation, since Eq.~(\ref{BasicGravCutting})
returns a tree amplitude multiplied by some scalar factors. The
three-particle cuts are more difficult to obtain, but again one can
``recycle'' the corresponding cuts used to obtain the two-loop $N=4$
super-Yang-Mills amplitudes~\cite{BRY}.  It turns out that the
three-particle cuts introduce no other functions than those already
detected in the two-particle cuts.  After all the cuts are combined into a
single function with the correct cuts in all channels, the $N=8$
supergravity two-loop amplitude~\cite{BDDPR} is:
\begin{eqnarray}
{\cal M}_4^{{\rm 2\,loop}}(1,2,3,4) & = &
  \Bigl({\kappa \over 2} \Bigr)^6 \!\! s_{12} s_{23} s_{13}
  M_4^{\rm tree}(1,2,3,4) \nonumber\\
&& \hskip .3 cm \times
\Bigl( s_{12}^2 \, {\cal I}_4^{{\rm 2\, loop},{\rm P}}(s_{12},s_{23}) 
+ s_{12}^2 \, {\cal I}_4^{{\rm 2\, loop},{\rm P}}(s_{12},s_{13}) \nonumber \\
& & \null \hskip  .7 cm  
+ s_{12}^2 \, {\cal I}_4^{{\rm 2\, loop},{\rm NP}}(s_{12},s_{23})
+ s_{12}^2 \, {\cal I}_4^{{\rm 2\, loop},{\rm NP}}(s_{12},s_{13}) \nonumber \\
& & \null \hskip  3 cm  
+\;  \hbox{cyclic} \Bigr) \,, 
\label{TwoLoopAmplitude}
\end{eqnarray}
where ``$+$~cyclic'' instructs one to add the two cyclic permutations of
legs (2,3,4). The scalar planar and non-planar loop momentum integrals, ${\cal
I}_4^{{\rm 2\,loop},{\rm P}}(s_{12}, s_{23})$ 
and ${\cal I}_4^{{\rm 2\,loop},{\rm
NP}}(s_{12}, s_{23})$, are depicted in 
Fig.~\ref{figure:PlanarNonPlanar}.  In this
expression, all powers of loop momentum have cancelled from the
numerator of each integrand in much the same way as at one loop,
leaving behind only the Feynman propagator denominators.  The explicit
values of the two-loop scalar integrals in terms of polylogarithms may
be found in Refs.~\cite{Smirnov99,Tausk99}.

%FIGURE
%%%%%%%%%%%%%%%%%%%%%%%%%%% 
%%%%%%%%%%%%%%%%%%%%%%%%%%

The two-loop amplitude (\ref{TwoLoopAmplitude}) has been used by
Green, Kwon, and Vanhove~\cite{GreenTwoLoop} to provide an explicit
demonstration of the non-trivial M-theory duality between $D=11$
supergravity and type II string theory.  In this case, the finite
parts of the supergravity amplitudes are important, particularly the
way they depend on the radii of compactified dimensions.

\begin{figure}[h]
  \def\epsfsize#1#2{0.5#1} \centerline{\epsfbox{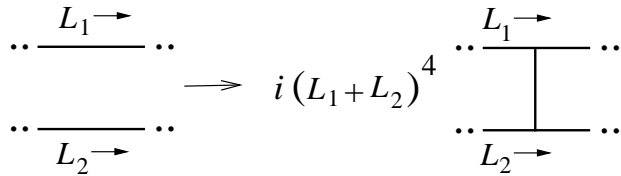}}
  \caption{\it Starting from an $l$-loop planar diagram representing
  an integral function, an extra line may be added to the inside using
  this rule.  The two lines on the left represent two lines in
  some $l$-loop diagram.}  \label{figure:InsertLine}
\end{figure}

A remarkable feature of the two-particle cutting equation
(\ref{BasicGravCutting}) is that it can be iterated to {\it all} loop
orders because the tree amplitude (times some scalar denominators)
reappears on the right-hand-side. Although this iteration is
insufficient to determine the complete multi-loop four-point
amplitudes, it does provide a wealth of information.  In particular,
for planar integrals it leads to the simple insertion rule depicted in
Fig.~\ref{figure:InsertLine} for obtaining the higher loop
contributions from lower loop ones~\cite{BDDPR}.  This class includes
the contribution in Fig.~\ref{figure:Multiloop}, because it can be
assembled entirely from two-particle cuts.  According to the insertion
rule, the contribution corresponding to Fig.~\ref{figure:Multiloop} is
given by loop integrals containing the propagators corresponding to
all the internal lines multiplied by a numerator factor containing 8
powers of loop momentum.  This is to be contrasted with the 24 
powers of loop momentum in the numerator expected when there are no
supersymmetric cancellations.  This reduction in powers of loop
momenta leads to improved divergence properties
described in the next subsection.

%FIGURE
%%%%%%%%%%%%%%%%%%%%%%%%%%% 
%%%%%%%%%%%%%%%%%%%%%%%%%%

%%%%%%%%%%%%%
\subsection{Divergence Properties of $N=8$ Supergravity}
\label{subsection:divergences}

Since the two-loop $N=8$ supergravity amplitude
(\ref{TwoLoopAmplitude}) has been expressed in terms of scalar
integrals, it is straightforward to extract the divergence properties.
The scalar integrals diverge only for dimension $D\ge 7$; hence the
two-loop $N=8$ amplitude is manifestly finite in $D=5$ and $6$,
contrary to earlier expectations based on superspace
power counting~\cite{Howe89}.  The discrepancy between the above
explicit results and the earlier superspace power counting arguments
may be understood in terms of an unaccounted higher dimensional gauge
symmetry~\cite{Stelle}. Once this symmetry is accounted for,
superspace power counting gives the same degree of divergence as the
explicit calculation.

The cutting methods provide much more than just an indication 
of divergence; one can extract the explicit numerical coefficients
of the divergences. For example, near $D=7$ the divergence of the
amplitude (\ref{TwoLoopAmplitude}) is:
\begin{equation}
{\cal M}_4^{{\rm two-loop},\ D=7-2 \epsilon}\Bigr\vert_{\rm div.} = 
 {1\over 2 \epsilon\ (4\pi)^7} {\pi\over3} (s_{12}^2+s_{23}^2+s_{13}^2) \, 
\left( {\kappa \over 2}\right)^6 
\times s_{12} s_{23} s_{13} M_4^{\rm tree} \,, 
\label{gravtwolooppoles}
\end{equation}
which clearly diverges when the dimensional regularization parameter 
$\epsilon \rightarrow 0$.

In all cases the linearized divergences take the form of derivatives
acting on a particular contraction of Riemann tensors, which in four
dimensions is equivalent to the square of the Bel-Robinson
tensor~\cite{Bel,Deser99BelRobinsonA,Deser99BelRobinsonB}.  This
operator appears in the first set of corrections to the $N=8$
supergravity Lagrangian, in the inverse string-tension expansion of
the effective field theory for the type II
superstring~\cite{GrossWitten}.  Therefore, it has a completion
into an $N=8$ supersymmetric multiplet of operators, even at the
non-linear level.  It also appears in the M-theory one-loop and
two-loop effective actions \cite{StringR4,Tseytlin,GreenTwoLoop}.

Interestingly, the manifest $D$-independence of the cutting algebra
allows the calculation to be extended to $D=11$, even though there is
no corresponding $D=11$ super-Yang-Mills theory. The result
(\ref{TwoLoopAmplitude}) then explicitly demonstrates that $N=1$
$D=11$ supergravity diverges.  In dimensional regularization there are
no one-loop divergences so the first potential divergence is at two
loops.  (In a momentum cutoff scheme the divergences actually begin at
one loop~\cite{Tseytlin}.)  Further work on the structure of the
$D=11$ two-loop divergences in dimensional regularization has been
carried out in Ref.~\cite{DeserSeminara99,DeserSeminara00}.  The
explicit form of the linearized $N=1$, $D=11$ counterterm expressed as
derivatives acting on Riemann tensors along with a more general
discussion of supergravity divergences may be found in
Ref.~\cite{Bern00Counterterms}.

Using the insertion rule of Fig~\ref{figure:InsertLine}, 
and counting the powers of loop momenta in these contributions 
leads to the simple finiteness condition:
\begin{equation}
l < {10 \over D-2}
\label{FinitenessFormuala}
\end{equation}
(with $l>1$), where $l$ is the number of loops.  This formula
indicates that $N=8$ supergravity is finite in some other cases where
the previous superspace bounds suggest divergences~\cite{Howe89}, {\it
e.g.} $D=4$, $l=3$: The first $D=4$ counterterm detected via the
two-particle cuts of four-point amplitudes occurs at five, not three
loops.  Further evidence that the finiteness formula is correct stems
from the maximally helicity violating contributions to $m$-particle
cuts, in which the same supersymmetry cancellations occur as for the
two-particle cuts~\cite{BDDPR}.  Moreover, a recent improved
superspace power count~\cite{Stelle}, taking into account a higher
dimensional gauge symmetry, is in agreement with the finiteness
formula~(\ref{FinitenessFormuala}).  Further work would be required to
prove that other contributions do not alter the two-particle cut power
counting.  A related open question is whether one can prove that the
five-loop $D=4$ divergence encountered in the two-particle cuts does not
somehow cancel against other contributions~\cite{ChalmersN8} because
of some additional symmetry.  It would also be interesting to
explicitly demonstrate the non-existence of divergences after
including all contributions to the three-loop amplitude.  In any case,
the explicit calculations using cutting methods do establish that at
two loops maximally supersymmetric supergravity does not diverge in
$D=5$~\cite{BDDPR}, contrary to earlier expectations from superspace
power counting~\cite{Howe89}.

\newpage

%=========================================================================
\section{Conclusions}
\label{section:conclusions}

This review described how the notion that gravity~$\sim$~(gauge
theory)~$\times$~(gauge theory) can be exploited to develop a better
understanding of perturbative quantum gravity.  The Kawai-Lewellen-Tye
(KLT) string theory relations~\cite{KLT} gives this notion a precise meaning 
at the semi-classical or tree level.  Quantum loop effects
may then be obtained by using $D$-dimensional
unitarity~\cite{Bern94SusyFour,Bern95SusyFour,BernMorgan,Review}.  In
a sense, this provides an alternative method for quantizing gravity,
at least in the context of perturbative expansions around flat space.
With this method, gauge theory tree amplitudes are converted into
gravity tree amplitudes which are then used to obtain gravity loop
amplitudes.  The ability to carry this out implies that gravity and
gauge theory are much more closely related than one might have deduced
by an inspection of the respective Lagrangians.

Some concrete applications were also described, including the
computation of the two-loop four-point amplitude in maximally
supersymmetric supergravity. The result of this and related
computations is that maximal supergravity is less divergent in the
ultraviolet than had previously been deduced from superspace power
counting arguments~\cite{BDDPR,Stelle}.  For the case of
$D=4$, maximal supergravity appears to diverge at five instead of
three loops.  Another example for which the relation is useful is for
understanding the behavior of gravitons as their momenta become either
soft or collinear with the momenta of other gravitons.  The soft
behavior was known long ago~\cite{WeinbergSoftG}, but the collinear
behavior is new.  The KLT relations provide a means for expressing the
graviton soft and collinear functions directly in terms of the
corresponding ones for gluons in quantum chromodynamics.  Using the
soft and collinear properties of gravitons, infinite sequences of
maximally helicity violating gravity amplitudes with a single quantum
loop were obtained by bootstrapping~\cite{AllPlusGrav,MHVGrav} from
the four-, five-, and six-point amplitudes obtained by direct
calculation using the unitarity method together with the KLT
relations.  Interestingly, for the case of identical helicity, the
sequences of amplitudes turn out to be the same as one gets from
self-dual gravity~\cite{Plebanski75,Duff79,Plebanski96}.

There are a number of interesting open questions.  Using the
relationship of gravity to gauge theory one should be able to
systematically re-examine the divergence structure of non-maximal
theories.  Some salient work in this direction may be found in
Ref.~\cite{DunbarJulia}, where the divergences of Type I supergravity
in $D=8, 10$ were shown to split into products of gauge theory
factors.  More generally, it should be possible to systematically
re-examine finiteness conditions order-by-order in the loop expansion
to more thoroughly understand the divergences and associated
non-renormalizability of quantum gravity.

An important outstanding problem is the lack of a direct derivation of
the KLT relations between gravity and gauge theory tree amplitudes
starting from their respective Lagrangians.  As yet, there is only a
partial understanding in terms of a ``left-right'' factorization of
space-time indices~\cite{Siegel93A,Siegel93B,Siegel94,BernGrant}, which
is a necessary condition for the KLT relations to hold.  A more
complete understanding may lead to a useful reformulation of gravity
where properties of gauge theories can be used to systematically
understand properties of gravity theories and vice versa.  Connected
with this is the question of whether the heuristic notion that gravity
is a product of gauge theories can be given meaning outside of
perturbation theory.

In summary, the perturbative relations between gravity and gauge
theory provide a new tool for understanding non-trivial properties of
quantum gravity. However, further work will be required to unravel
fully the intriguing relationship between the two theories.

%=========================================================================

\section{Acknowledgments}
\label{section:acknowledgements}

The author thanks Abilio De Freitas, Aaron Grant, David Dunbar, David
Kosower, Maxim Perelstein, Joel Rozowsky, Henry Wong, and especially
Lance Dixon for collaboration on work described here and for sharing
their insight into quantum gravity.  The author also thanks Eduardo
Guendelman for a number of interesting discussions on the Einstein-Hilbert
action and its relation to gauge theory.  This work was supported by
the US Department of Energy under grant
DE-FG03-91ER40662.

\newpage

%========================================================================

%\bibliography{LivRevTemplate}

\end{document}